\documentclass[sigconf,10pt]{acmart}

\usepackage[english]{babel}
\usepackage{blindtext}
\usepackage{epsfig,endnotes}
\usepackage{graphicx}
\usepackage{amsmath}
\usepackage{float}
\usepackage[textfont=it]{caption}
\usepackage{wrapfig}
\usepackage{multirow, multicol, booktabs}
\usepackage[labelfont=sc]{caption}
\usepackage{graphbox}
\usepackage{lipsum}
\usepackage[linesnumbered,algoruled,boxed,lined]{algorithm2e}
\usepackage{algpseudocode}
\usepackage[export]{adjustbox}
\usepackage{subfig}
\usepackage{bm}
\usepackage{booktabs}
\usepackage{tabularx}
\usepackage{pifont}
\usepackage{titlesec}
\usepackage{xcolor}
\usepackage{hyperref}
\hypersetup{colorlinks=true, linkcolor=blue, urlcolor=blue}
\usepackage{url}
\usepackage{breakurl}

\DeclareMathOperator{\sinc}{sinc}

\definecolor{Cerulean}{HTML}{007FFF}

\newcommand{\name}{BYON}
\newcommand{\para}[1]{\vspace{1pt}\noindent\textbf{#1}}
\newcommand{\squishlist}
{
    \begin{list}{$\bullet$}
    {
        \setlength{\itemsep}{0pt}      \setlength{\parsep}{3pt}
        \setlength{\topsep}{3pt}       \setlength{\partopsep}{0pt}
        \setlength{\leftmargin}{1.5em} \setlength{\labelwidth}{1em}
        \setlength{\labelsep}{0.5em}
    }
}
\newcommand{\squishend}
{
    \end{list}
}

\renewcommand\footnotetextcopyrightpermission[1]{} %
\setcopyright{none}

\acmDOI{}

\acmISBN{}

\acmPrice{}

\titlespacing*{\section}{2pt}{0pt}{0pt}
\titlespacing*{\subsection}{2pt}{0pt}{0pt}

\begin{document}

\title[\name: Bring Your Own Networks]{\LARGE \bf \name: Bring Your Own Networks for Digital Agriculture Applications}

\author{\large Emerson Sie, Bill Tao, Aganze Mihigo, Pari Karmehan, Max Zhang, Arun N. Sivakumar, \\ Girish Chowdhary, Deepak Vasisht \\ \vspace{0.1in} University of Illinois Urbana-Champaign}

\renewcommand{\shortauthors}{X.et al.}

\begin{abstract}
Digital agriculture technologies rely on sensors, drones, robots, and autonomous farm equipment to improve farm yields and incorporate sustainability practices. However, the adoption of such technologies is severely limited by the lack of broadband connectivity in rural areas. We argue that farming applications do not require permanent always-on connectivity. Instead, farming activity and digital agriculture applications follow seasonal rhythms of agriculture. Therefore, the need for connectivity is highly localized in time and space. We introduce BYON, a new connectivity model for high bandwidth agricultural applications that relies on emerging connectivity solutions like citizens broadband radio service (CBRS) and satellite networks. BYON creates an agile connectivity solution that can be moved along a farm to create spatio-temporal connectivity bubbles. BYON incorporates a new gateway design that reacts to the presence of crops and optimizes coverage in agricultural settings.  We evaluate BYON in a production farm and demonstrate its benefits.\vspace{-0.2in}
\end{abstract}

\maketitle
\pagestyle{plain}

\section{Introduction}
\label{sec:intro}

Digital agriculture technologies, such as precision agriculture, agricultural robotics (AgBots), and the Internet of Things (IoT), can improve agricultural outcomes by tens of billions of dollars annually in the United States \cite{usda_report, fcc_report, mckinsey_agriculture_2020} and around the globe \cite{mehrabi_global_2021, 30_percent_no_internet, govuk_ruraldrive_2023}. These technologies are also key to improving sustainability outcomes in agriculture. For example, agricultural robots can plant cover crops before harvest and prevent soil erosion, reduce fertilizer use, and replace chemical herbicides with mechanical weeding \cite{digital_transformation_madhu, farmprogress_robot, mckinsey_agriculture_2020}. In practice, the adoption of such technologies is severely limited by the lack of broadband connectivity \cite{30_percent_no_internet, usda_report, comparing_on-farm_connectivity, mehrabi_global_2021}.

High-bandwidth connectivity is essential to stream data from farm equipment, to teleoperate robots, and to operate data-intensive farm equipment like autonomous tractors and drones~\cite{johndeere,johndeere2,mckinsey_agriculture_2020}. However, existing connectivity solutions fail to meet these needs. Wi-Fi operates in short ranges from fifty to hundred meters, thereby failing to cover large farmlands spanning several Kilometers. Narrowband connectivity solutions like LoRa enable few Kbps of connectivity for sensors, but cannot support cameras, robots, or tractors. Commercial cellular operators often don't cover farm areas due to their sparse population and high deployment costs~\cite{usb_report, 30_percent_no_internet}. Different estimates argue that 30-50\% of the farmers in the US have limited or no access to broadband connectivity on their farms~\cite{usb_report, 30_percent_no_internet}. Therefore, there is a pressing need for high-bandwidth, low-cost, and long-range connectivity solutions for digital agriculture applications. %

\begin{figure}[t]
    \centering
    \includegraphics[width=0.65\linewidth]{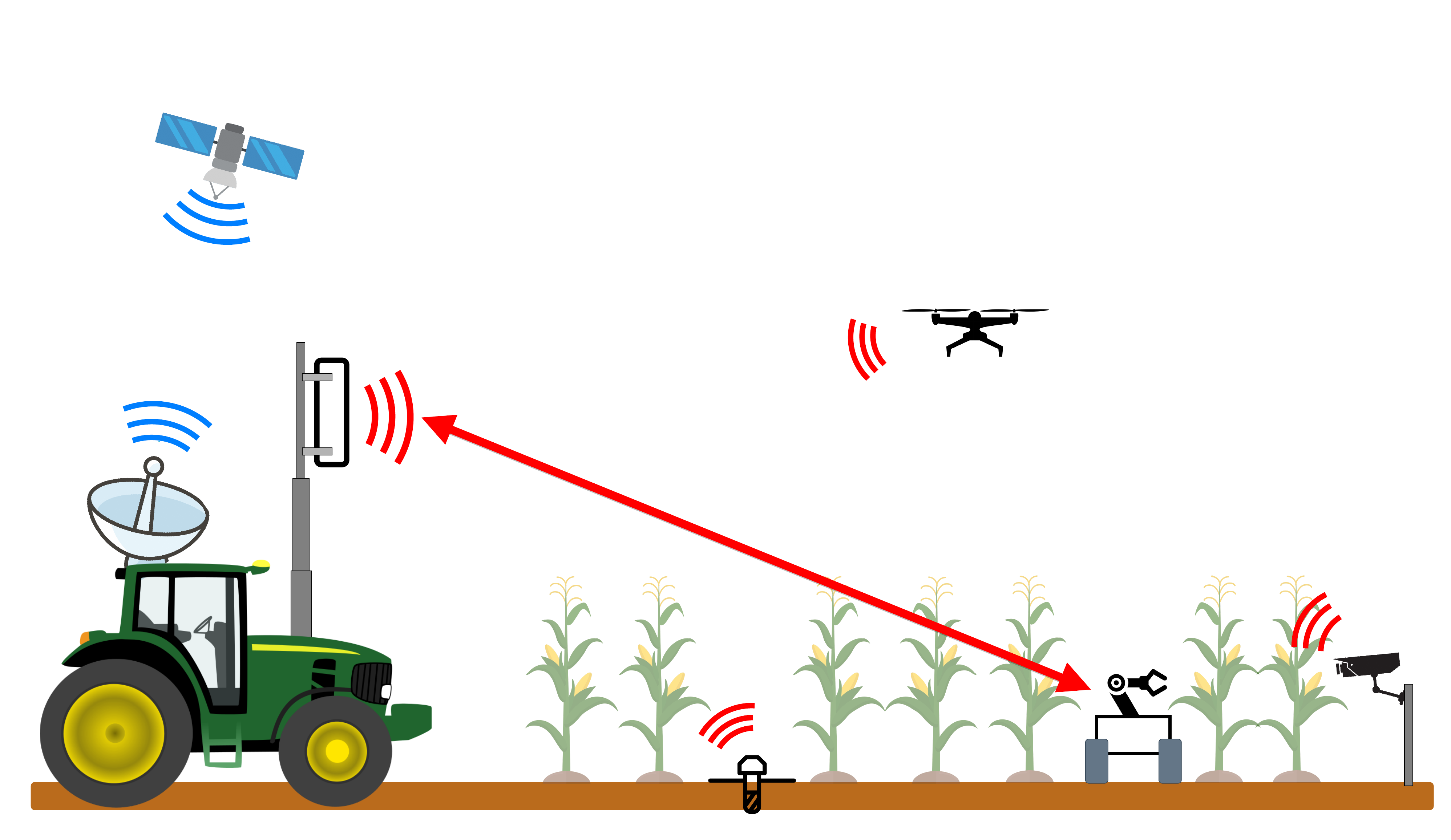}
    \vspace{-0.1in}
    \caption{BYON gateway consisting of CBRS base station, telescoping antenna mast, and satellite terminal.}%
    \label{fig:byon}
    \vspace{-0.20in}
\end{figure}

In this paper, we argue that the goal of providing always-on connectivity that connects all of the farmland is an over-estimation of the farm connectivity problem. Instead, we observe that the demand for agricultural connectivity is restricted in time and space, i.e., agricultural equipment typically needs coverage in small parts of the farm and this requirement varies with time much like the seasonal rhythm of agricultural activity. Agricultural equipment (increasingly autonomous and video-based) sequentially plants crops at different locations in a farm across few days. Agricultural robots for mechanical weeding operate in a small window before and after that. Other robots plant cover crops in a short window of time before crop harvest. Therefore, such applications do not require permanent and full-coverage high-bandwidth connectivity. In fact, efforts to provide such connectivity, e.g., using cellular networks with dense backhauls, end up being cost-prohibitive and severely over-provisioned.

As opposed to permanent and full-coverage solutions, we seek to enable a new networking model where the farm network moves across space and time, along with the applications themselves. Specifically, \textit{we propose BYON (Bring Your Own Network) -- a new connectivity model} where high-bandwidth applications bring their own network to different parts of the farm as needed. %
The design of BYON leverages two emerging technologies. (a) We use Citizens Broadband Radio Service (CBRS)~\cite{cbrs}, which promises to enable citizen-deployed private cellular networks using shared spectrum, to provide last-mile connectivity across various edge devices. (b) BYON leverages recent satellite connectivity solutions \cite{starlink} to provide a globally accessible backhaul to the Internet. 

BYON offers multiple advantages: (a) BYON solutions are  portable and mobile. For example, it can be deployed on a tractor (see Fig.~\ref{fig:byon}) that parks at the edge of the field being worked on in a given day and moves to different fields in a farm over time. (b) BYON is compatible with off-the-shelf cellular devices, i.e., it simply requires a new SIM card to be installed in cellular devices. A USB dongle can provide cellular capabilities to devices like robots. So, BYON can provide connectivity to both farm devices and farm workers who work the field. (c) CBRS offers long range connectivity as opposed to Wi-Fi and can support operations over a large part of the farm without requiring additional movement. (d) BYON is highly configurable and can meet the need for both over-canopy and under-canopy applications.

In this paper, we focus on three key aspects of BYON:

\para{Profiling CBRS Connectivity: } CBRS is an emerging form of private cellular networks, wherein anyone can utilize the shared spectrum (up to 150 MHz) recently opened by FCC to deploy citizen-driven cellular networks~\cite{cbrs}. This is particularly targeted towards rural areas, where existing cellular networks do not have sufficient coverage. We deploy an off-the-shelf CBRS network on a production agricultural farm and profile its performance. In a 20 MHz band, we find that CBRS achieves downlink bandwidth up to 100 Mbps, and uplink bandwidths up to 20 Mbps across a distance of up to 4 Kilometers. However, due to its high frequencies around 3.5 GHz, such links suffer additional attenuation up to 30 dB for under-canopy applications. The additional attenuation leads to significantly reduced datarates under crop-canopy. %

\para{Variable Height Base Station Deployments: }As mentioned above, providing connectivity to sensors and robots covered by crops is a significant challenge. For digital agriculture applications, under-canopy robots and sensors play a crucial role. However, unlike traditional cellular infrastructure, BYON serves a small set of applications and can be dynamically configured to meet the needs of these applications. We propose a new base station design that can adapt its own height  based on crop-levels and application demand. 

In free space, the signal quality between the base station and a client device depends on the distance, $d$, between them (typically as $\frac{1}{d^2}$). However, as crops grow, the signal quality also degrades with the distance travelled through crops. Unlike free space, the attenuation  through crops is exponential with distance $e^{-\alpha d}$, where $\alpha$ depends on the electrical permittivity of the crops. This attenuation leads to an interesting tradeoff. If we set the base station height too low, the signal from the base station to the client travels a near-horizontal path which minimizes the distance and hence, the free-space attenuation. However, it travels a large distance through crops and maximizes the exponential attenuation through crops. On the other hand, at higher heights, the distance between the base station and the client increases, but the distance through crops decreases. We build a model to balance these competing factors and compute the optimal height. Our model computes the ideal height based on the base station location, throughput requirements, and crop heights. The height of the base station can be varied dynamically using computer-controlled telescoping antenna masts \cite{aluma_smarttower, willburt_mast_stilleto}, and can be controlled to serve different applications or different locations at different times of the day.

\para{Application Analysis -- Under-Canopy Robot Teleoperation: }We demonstrate the benefits of our connectivity solutions in a robot teleoperation application. Under-canopy robots are increasingly used in agriculture for applications like plant phenotyping~\cite{phenotyping1,sivakumar_learned_2021,manish_agbug_2021,kim_p-agbot_2022}, cover crop planting~\cite{icover_usda, farmprogress_robot, du_deep-cnn_2022} and mechanical weeding \cite{naio_oz, mcallister_agbots_2020, reiser_development_2019}. They hold major advantages over robots featuring elements extending above the crop canopy (e.g. \cite{mineral_rover,xiang_fieldbased_2023,xu_modular_2022}) as they (a) do not make contact with the canopy while following crop rows, which damages crop leaves and contributes unwanted drag forces onto the robot, and (b) can easily access key parts of the crop underneath the canopy including their stems and the soil, where weeds, pests and disease are likely to reside. %
In practice, the lack of connectivity due to crop canopy blockage is a key bottleneck for widespread adoption of these robots. For example, state-of-the-art under-canopy robots require manual intervention when they get stuck~\cite{sivakumar_learned_2021, velasquez_multi-sensor_2022,gasparino_cropnav_2023}. Due to the lack of connectivity, a human manually needs to walk to these robots in the field and maneuver them when they get stuck and cannot decide. Second, such robots cannot exchange information with other robots when they work in groups. We demonstrate long range teleoperation for such robots, which would not be possible with existing techniques like Wi-Fi or TV White Spaces. In a BYON setup, a human sees a video feed while being away from the robot when the robot gets stuck and can remotely teleoperate the robot through thick crop canopy cover without requiring physical intervention.

We perform measurements over farmland at distances of upto 3.6 Km. We deployed our solutions and evaluated them on a production farm. Our experiments establish the feasibility of CBRS for high-bandwidth agricultural applications, but demonstrate the challenges related to under-canopy coverage. We demonstrate that our height-variable base station design can increase median signal quality by 7.5 dBm and median throughput by 28\% for under-canopy applications. Finally, we demonstrate BYON's feasibility for under-canopy teleoperation. A demo video is available at \href{https://byon-v1.github.io/}{\color{blue}https://byon-v1.github.io/}.

Our work is novel in the underlying technologies (CBRS and satellite networks), new system design, and the application. Specifically, we make the following contributions:
\squishlist
    \item We present the first analysis of a CBRS network in a digital agriculture setup and quantify the impact of crops on the network performance. 
    \item We propose an agile variable height base station design for optimizing under-canopy coverage.
    \item We demonstrate the first teleoperation operation for an under-canopy operation over CBRS.
\squishend

\section{Background and Related Work}\label{sec:background}

Digital agriculture spans a range of applications aimed at enhancing productivity, reducing input costs for farmers, and reducing environmental harms. Increasingly, such techniques rely on a combination of sensors, drones, robots, and farm equipment~\cite{farmbeats}. For example, agriculture has relied on herbicide-based weeding for a long time due to the efficiency of herbicide application. However, herbicide application leads to chemically-resistant weeds and have other well-documented side-effects. Recent research develops small robots that can identify and remove weeds mechanically \cite{weeding_bots, weeding_bots2, naio_oz, reiser_development_2019}. Such robots reduce chemical use, and also reduce soil compaction caused by the weight of large equipment~\cite{uyeh2021evolutionary,van2022glyphosate}. Even traditional equipment like smart tractors and combines are increasingly equipped with sensors and cameras to collect extensive data about a farm for precision agriculture tasks. Finally, farm workers increasingly rely on connectivity for sharing farm images with each other or to access farm productivity software/services.

To serve most digital agriculture applications, we must consider two layers of connectivity: (a) connectivity \textit{to} a farm (e.g, Internet fiber) that allows farmers and equipment to download information from the Internet, upload data, update software, etc.; and (b) connectivity within the farm (e.g., Wi-Fi, LoRa, TV White Spaces) that enables farmers to provide feedback to autonomous equipment (e.g., does this picture contain a weed), perform teleoperation (e.g., when a robot gets stuck), and collect data for centralized processing. Today applications requiring broadband connectivity suffer from lack of both types of connectivity.

\para{Connectivity to the Farm: }Most broadband monitoring and deployment efforts, both government and commercial, focus on connecting people. Farms are sparsely populated. Therefore, companies have little incentive to deploy traditional connectivity infrastructure like fiber or cellular connectivity. For example, laying fiber can cost up to 10,000 US dollars per mile~\cite{ifn_cost}. This challenge is further exacerbated by the large span of farmlands (e.g., one-third of United States is farmlands~\cite{agcensus}). Therefore, it is infeasible to provide always-on and complete coverage to farms using traditional infrastructure-heavy solutions~\cite{ifn_cost,usb_report}. 

\para{Connectivity within the Farm: }Connecting to robots, smart tractors, and sensors on the field requires us to extend connectivity from the edge of the field (e.g., fiber connectivity that goes to a farm shed or farmer's home) to these devices. This is challenging because of: (a) Range: farms span a few miles. (b) Terrain: Local hills and valleys cause problems for signal propagation. (c) Crops: Crops block radio signals due to their high water content~\cite{wu2017propagation}. Technologies like Wi-Fi do not meet the range and coverage, while LoRa fails to achieve high bandwidth.

\subsection{Related Work}
\noindent\textbf{Digital Agriculture: }There has been much recent work in digital agriculture in the networking community spanning new connectivity solutions such as Whisper~\cite{whisper} and FarmBeats~\cite{farmbeats}, and new sensing solutions such as Strobe~\cite{strobe}, smol~\cite{smol}, GreenTag~\cite{greentag}, and Comet~\cite{comet}. FarmBeats focuses on leveraging a combination of sensors and drone imagery to extract insights about the farm. FarmBeats relies on TV White Spaces (TVWS) for to-farm connectivity and Wi-Fi for within farm connectivity. Wi-Fi has limited range and therefore, FarmBeats proposes deploying multiple gateways on the farm. In contrast, \name\ utilizes CBRS for on-farm connectivity and does not require multiple gateways. Instead, a single gateway moves to different parts of the farm to provide connectivity. Unlike FarmBeats, our work also profiles (and counters) the effect of crops on CBRS signals.

Similarly, Whisper uses TVWS for narrow-band transmissions for on-farm connectivity, but does not support high bandwidth applications. In general, TV White Spaces (TVWS)~\cite{whitefi,whisper} is a valuable connectivity platform for rural applications due to the availability of empty TV spectrum. TVWS can offer tens of Mbps of data rate and provide long range. However, TVWS require custom hardware and large antennas (due to their low frequencies) that can't be easily carried by drones or under-canopy robots. BYON models proposed in this paper can incorporate TVWS-based hardware where such hardware is available and feasible. 

Our work is also orthogonal to radio-based sensing ~\cite{strobe,smol,greentag,comet} as we do not seek to sense soil health based on radio signal propagation. Instead, we adapt the base station deployment to respond to crop-induced throughput degradations.

Finally,~\cite{wu2017propagation} has studied crop-induced attenuation for Wi-Fi signals, but we are the first to study the impact on CBRS spectrum. In addition, unlike~\cite{wu2017propagation}, we identify mechanisms to deal with such attenuation.

\para{Mobile Base Station Deployments: }There has been recent work on deploying cellular base stations on mobile platforms, often referred to as Cells-on-Wheels (CoWs). CoWs are designed to respond to increased traffic demands (e.g., in stadiums), disaster situations, or public safety use cases~\cite{cows1,911now}. Our work, \name, belongs to this category of research, but expands this along three axes: (a) we focus our work on the agriculture context where crops and the seasonal variation of crops plays an important role in signal quality; (b) while CoWs are generally agnostic to short-term traffic variation, \name\ explicitly adapts to application requirements by adapting the height and orientation of the base station; (c) we conduct an end to end study of a teleoperation use case in under-canopy robots. 

\name\ is also related to work in drone-based cellular deployments~\cite{drone_lte_1,dronelte2,dronelte3, kalantari2016number}. Such networks can vary the base station position depending on application demands. We believe drone-based deployments are harder to mount due to the battery constraints or the cost of deployment in tethered drone operation. However, where such deployments are feasible, they can benefit from insights in \name\ about how to adapt to different crop patterns, heights, etc.

Finally, past work such as ~\cite{antennaradiation,antennatilt,kovsmerl2014base} has extensively studied optimal placement of cellular base stations, setting the right orientation of antennas, and improving coverage. However, in such cases, the base station placement and configuration optimizes for blanket coverage. In contrast, \name\ optimizes for targeted coverage in agricultural scenarios and can benefit from knowledge of the application. For example, a \name\ setup can use the knowledge of a robot's path to have a coverage bubble follow the robot by changing its height and orientation, while a general base station placement algorithm does not have access to such application-level information.

\section{Our Approach: Bring your Own Networks}
\label{sec:approach}

\begin{figure*}[!ht] %
  \centering
  \subfloat[Under-canopy robot with CBRS dongle.]{%
    \includegraphics[width=0.30\textwidth, valign=c]{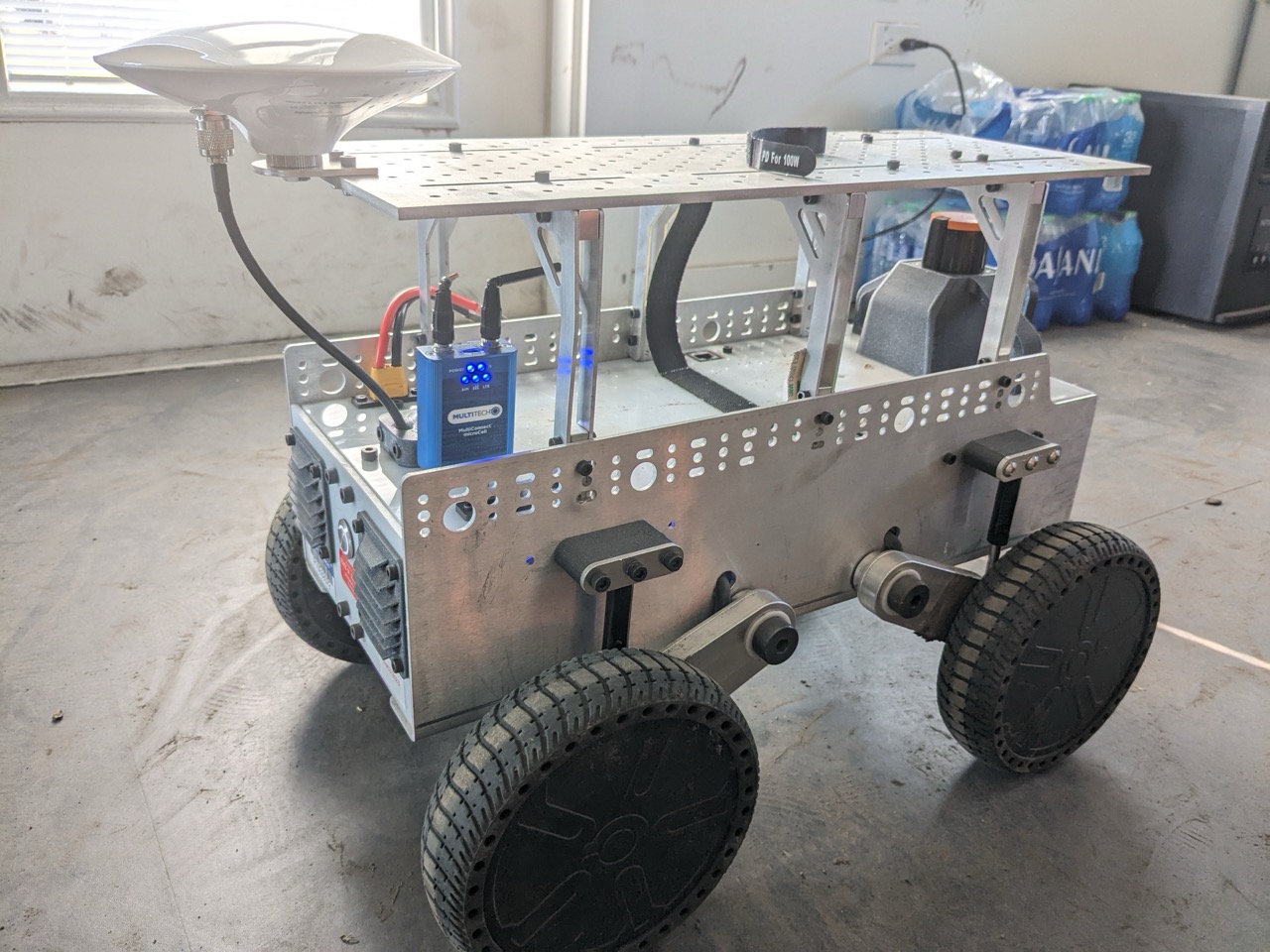} %
    \label{fig:robot}
  }\hfill
  \subfloat[In crops.]{%
    \includegraphics[width=0.17\textwidth, valign=c]{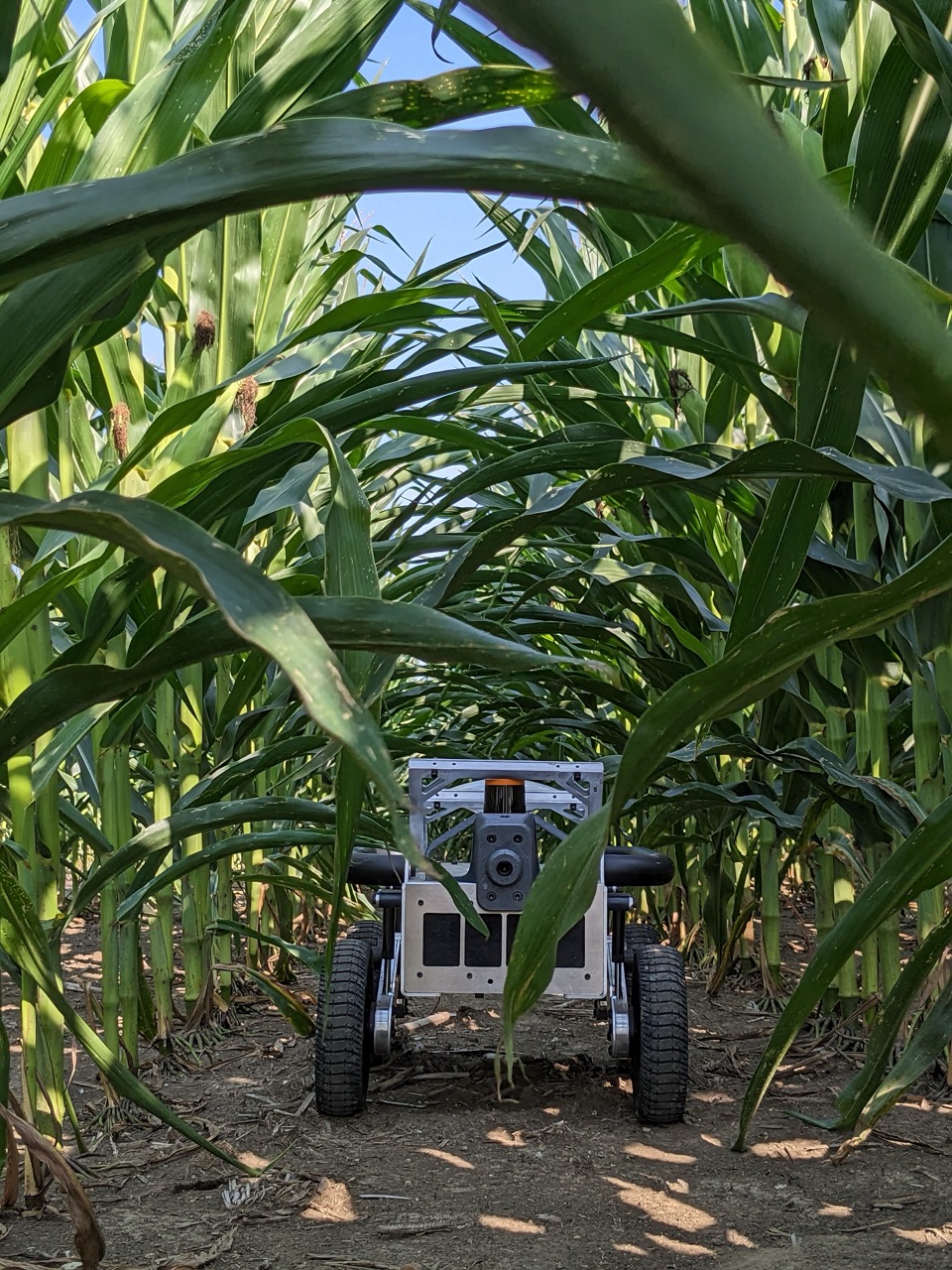} %
    \label{fig:crop-canopy}
  }\hfill
  \subfloat[Alternate view.]{%
    \includegraphics[width=0.17\textwidth, valign=c]{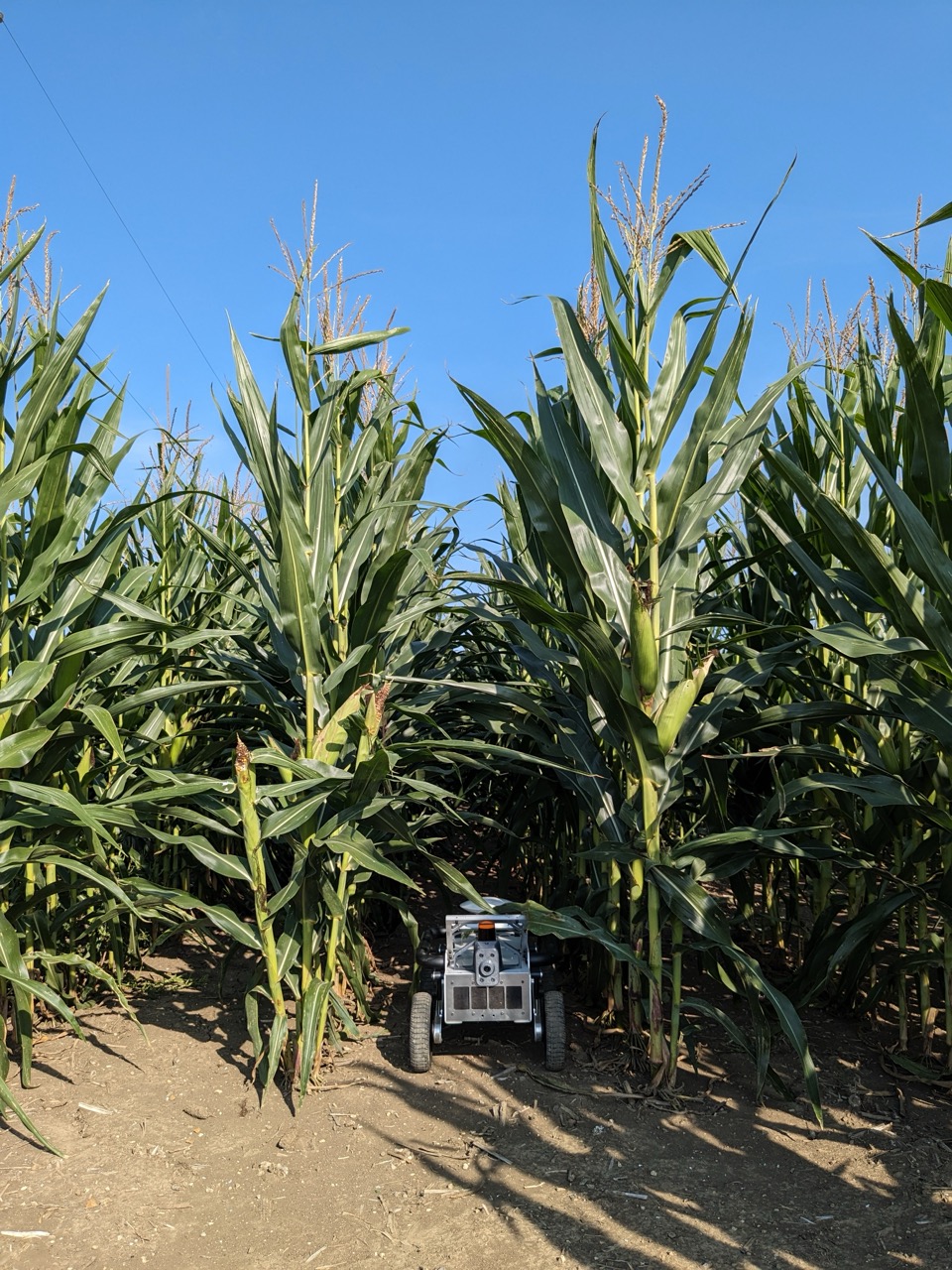} %
  }\hfill
  \subfloat[Plant phenotyping route (yellow). Crop rows in green.]{%
    \includegraphics[width=0.18\textwidth, valign=c]{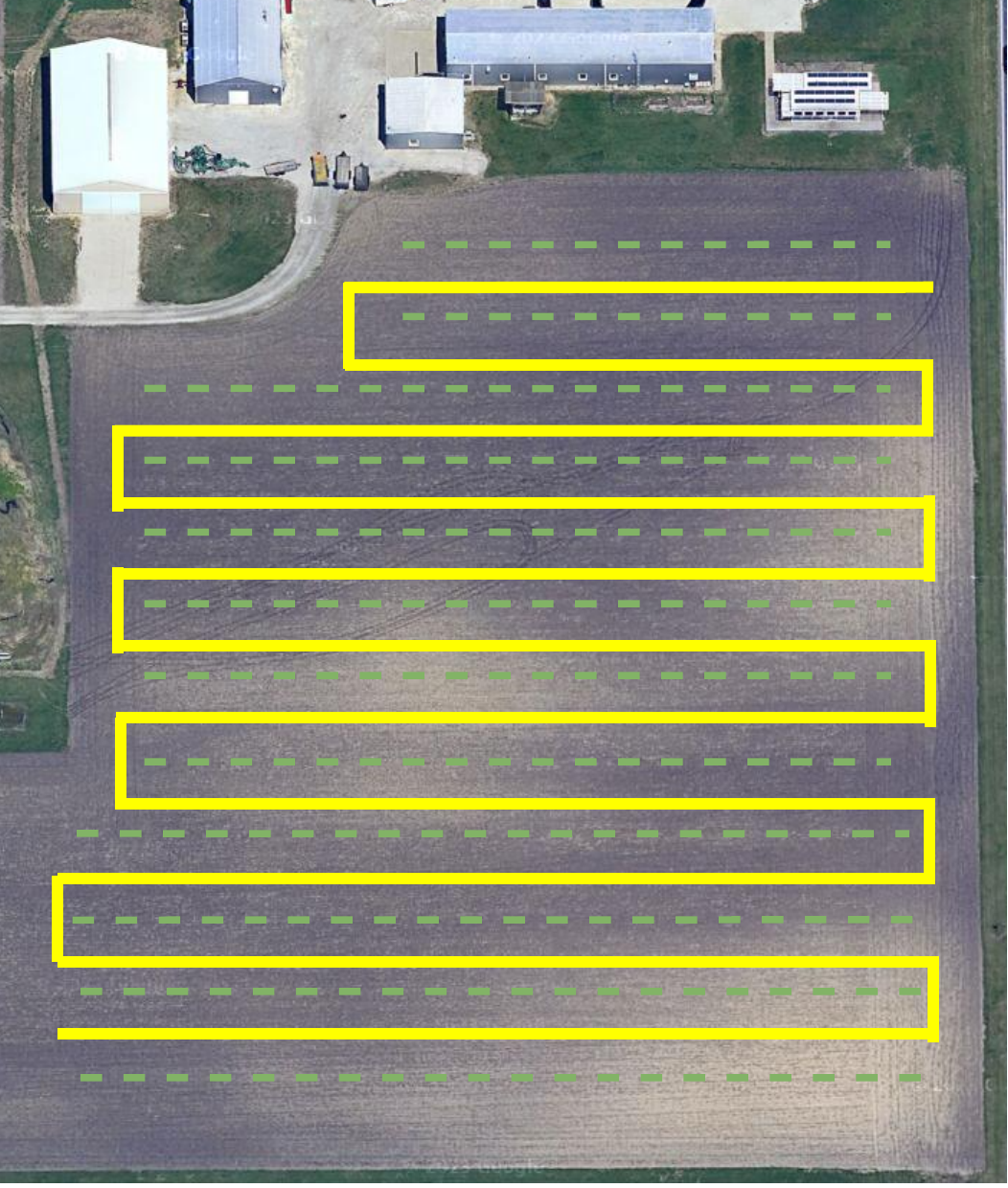} %
    \label{fig:phenotyping_route}
  }
  \vspace{-0.15in}
  \caption{\textbf{Under-canopy robots.} Robots are 40cm tall. In peak season, crop heights can reach 2m or more. }
  \label{fig:under-canopy}
  \vspace{-0.2in}
\end{figure*}

We propose a BYON~(Bring Your Own Network) connectivity model for digital agriculture applications. In BYON, farmers or farming communities do not need to sustain a permanent always-on network deployment that incurs large deployment costs. BYON relies on a portable deployment on a farm tractor or truck with a CBRS~(Citizens Broadband Radio Service) base stations with a satellite networking service (e.g., Starlink) as a backhaul. We choose CBRS because it allows users to deploy a private cellular network that offers mile-level range. CBRS spectrum has been recently opened up by regulators to target rural areas, where much of this spectrum is available. 80 MHz of this spectrum is always available to general users. This can go up to 150 MHz if priority users aren't using it at a given location~\cite{cbrs}. CBRS is compatible with off-the-shelf devices using a cellular SIM card. In devices without cellular capability, it can be attached using a cheap USB dongle. CBRS also offers a mile-level range.

We considered multiple choices for the backhaul -- fiber, cellular backhaul, satellite backhaul, and microwave backhaul. Fiber and cellular backhauls may not always be available in rural areas. Microwave backhauls require careful alignment every time the base station is moved. We decided to choose satellite backhauls because their performance has steadily improved and generally surpasses the requirements of our CBRS network. Finally, there has been recent push to incorporate satellite connectivity into tractors~\cite{deerespacex} which naturally aligns with the BYON model. We note that we cannot place satellite transceivers directly on end devices because  %
satellite receivers are typically bulky for drones and robots. More importantly, satellite transceivers use high frequencies which cannot penetrate cover-canopy for under-canopy applications.

\para{Target Applications:} Through this paper, we focus on video streaming from farm workers or farm equipment such as robots, tractors, or drones as the target application. Video is the highest bandwidth requirement for such equipment. Increasingly, such equipment is commonly used in agriculture. These applications regularly need to stream videos to a farm worker to seek feedback, decide on inputs (e.g., whether to kill a weed), and for help when they get stuck. Farm workers may also need to share images of what they observe on the field with other workers or the farm manager. Such applications can't be satisfied by low bandwidth technologies like LoRa. An example of such robots is shown in Fig.~\ref{fig:robot}. %

We note that there are some applications in digital agriculture like sensor-based monitoring which require permanent connectivity. However, such applications are usually low bandwidth and can be supported by low power wide area connectivity solutions (LPWANs) like LoRa or NB-IoT and are not the focus of our work. We instead focus on high-bandwidth dynamic applications like farm robots, drones, tractors, etc.

\section{Measurement Study}
\label{sec:measurements}

\begin{figure}[ht!] %
  \centering
  \subfloat[Cell tower.]{%
    \includegraphics[width=0.21\linewidth]{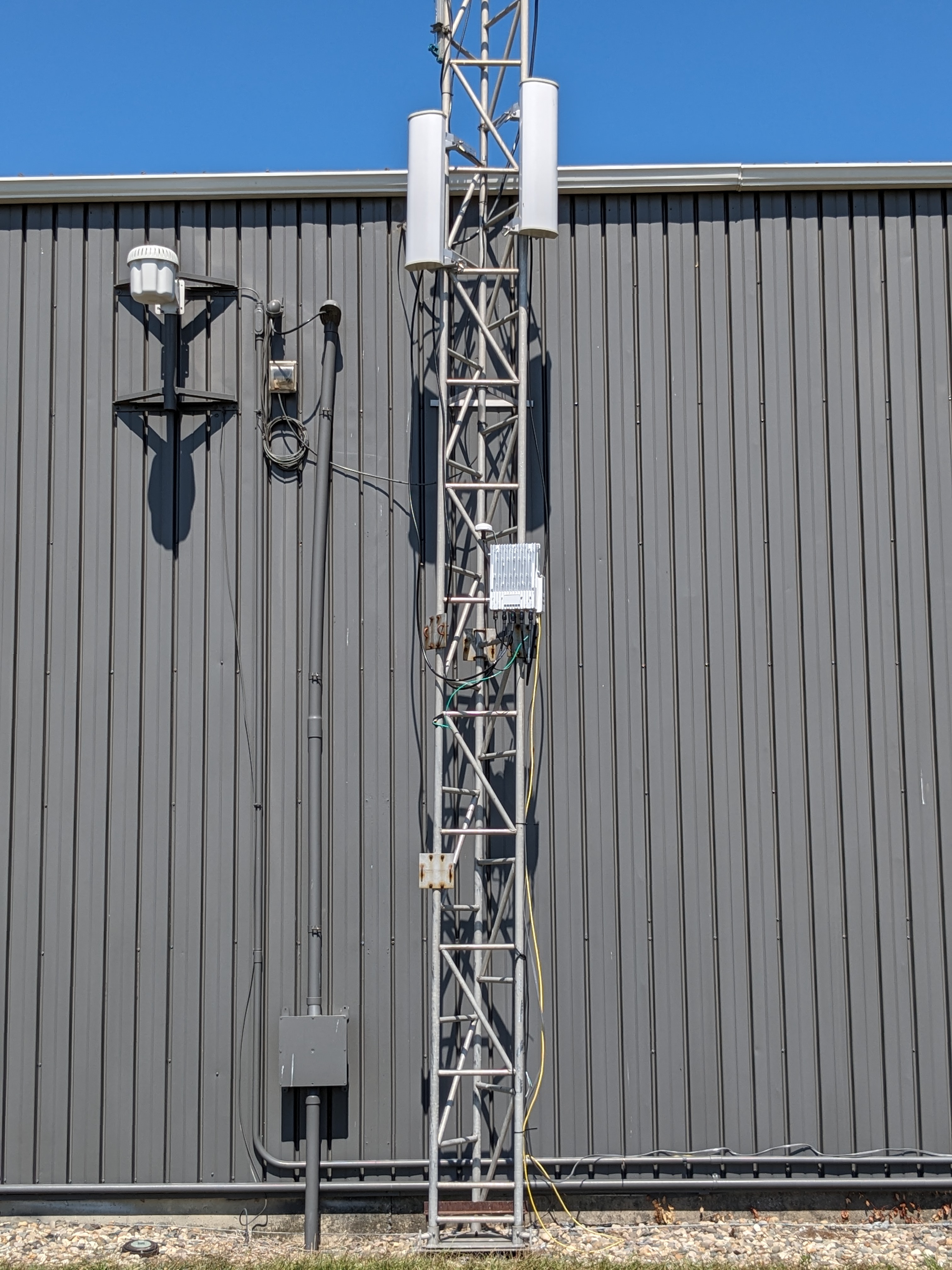} 
    \label{fig:base_station}
  }
  \subfloat[Overlooking crops.]{%
    \includegraphics[width=0.375\linewidth]{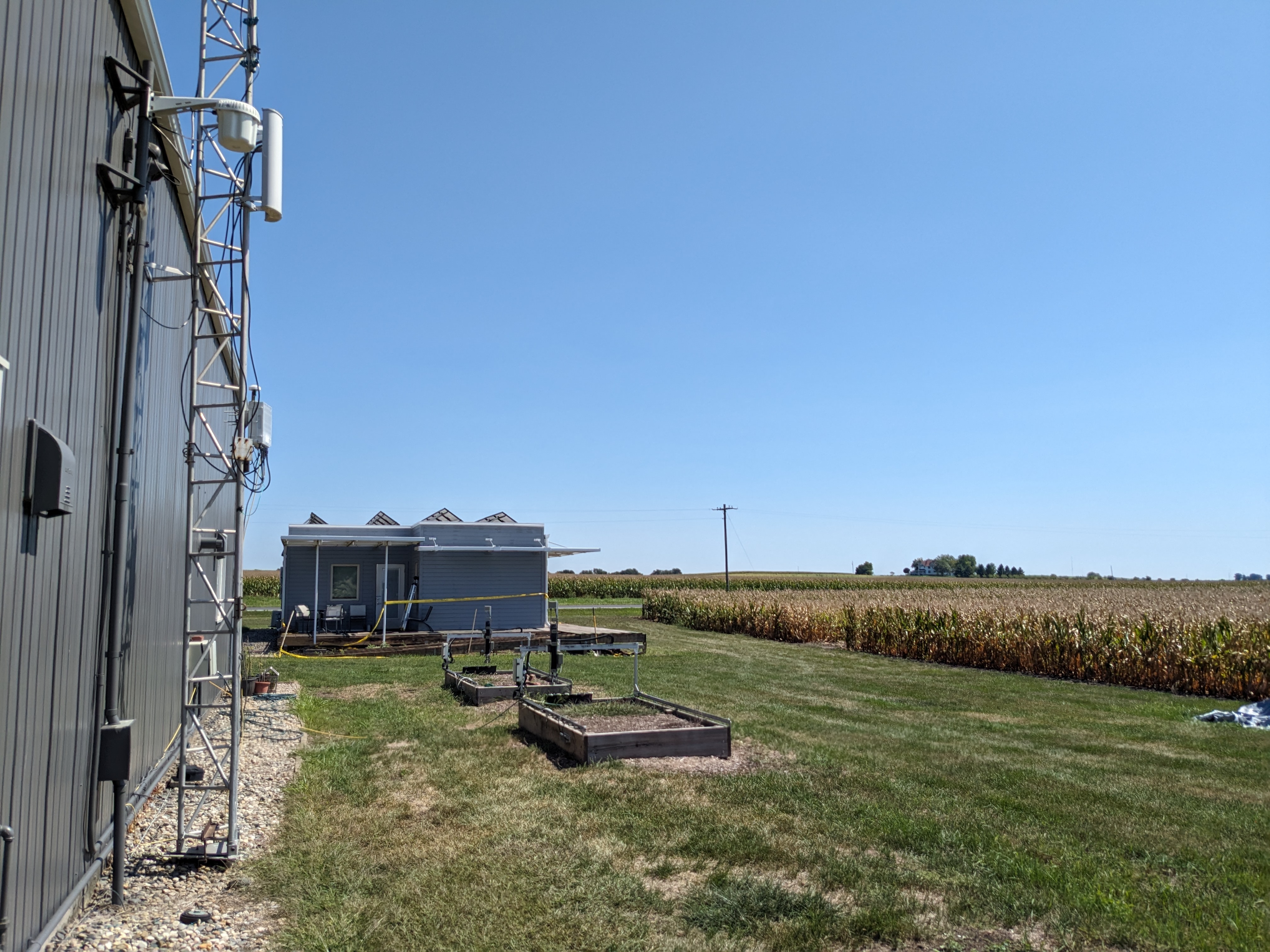} 
  }
  \subfloat[BYON deployment on a movable shipping container.]{%
    \includegraphics[width=0.375\linewidth]{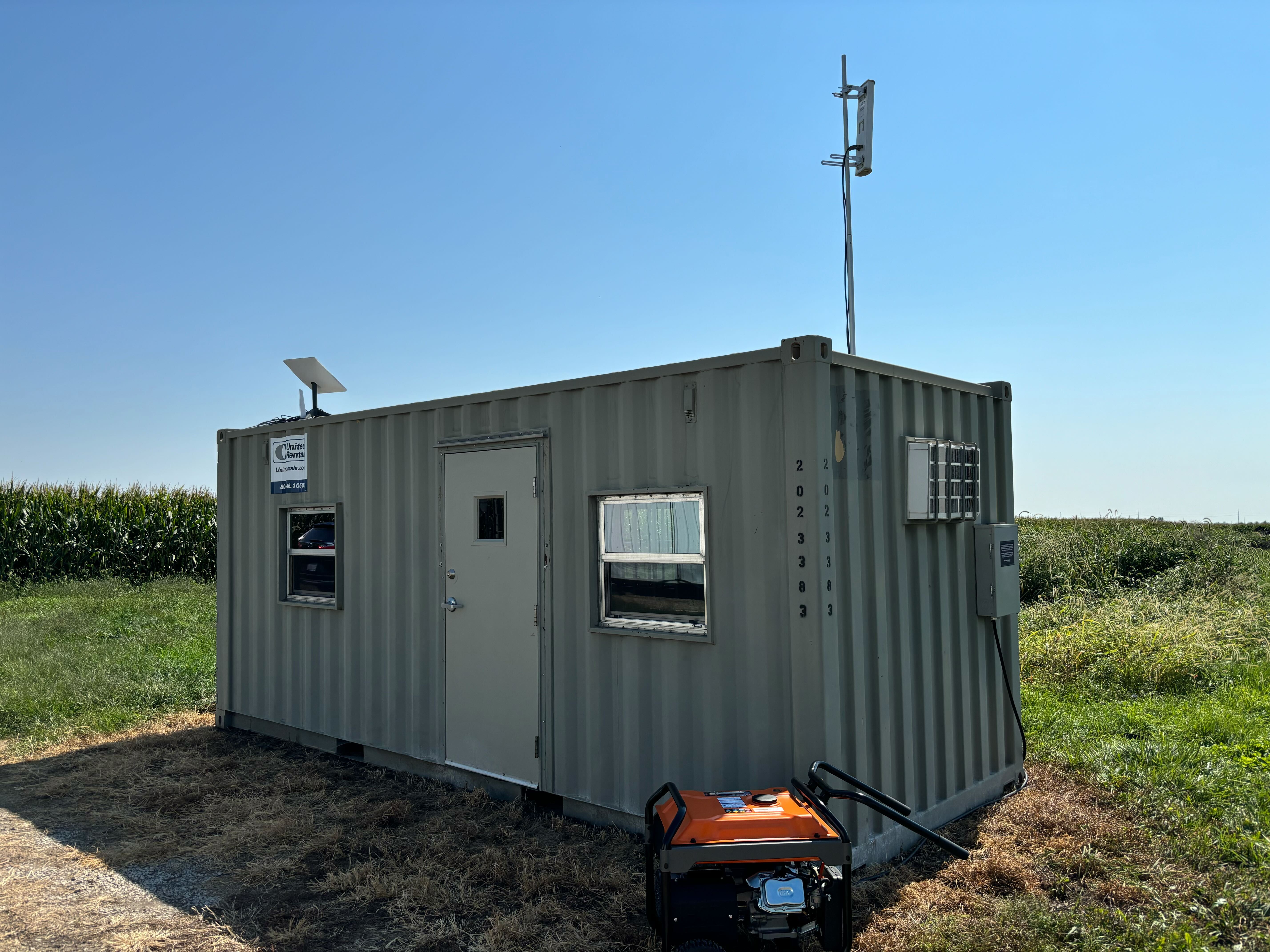} 
  }
  \vspace{-0.15in}
  \caption{\textbf{CBRS deployments on the farm.} (A,B) Conventional deployment on static infrastructure. (C) BYON is a containerized solution that can be moved on demand. } 
  \label{fig:deployment}
  \vspace{-0.25in}
\end{figure}

In this section, we characterize the performance of a production-grade CBRS network and a satellite terminal deployed in a corn-field. We present an overview of the range, throughput, and latency that client devices on the farm experience under real world conditions.

\para{Deployment:} Our CBRS network deployment is pictured in Fig.~\ref{fig:deployment}. We mount a Celona AP11 Outdoor Access Point (AP) \cite{celona_ap11} onto a small  tower. The AP is connected to two 90-degree sectorized slant dual-polarization antennas (CN-ANT-90D \cite{celona_antenna}). The access point is set to operate in the CBRS bands in United States around 3.5 GHz, with a 20 MHz bandwidth. The antennas are mounted onto the tower at a height of 5m and directed to point southward. For maximum coverage, we use the maximum transmit power setting of 50W. We connect clients to the CBRS network using the Multi-Tech microCell CBRS dongle \cite{multitech_multiconnect} pictured in Fig.~\ref{fig:robot}. The dongle is lightweight and connects to robots, drones, sensors, etc. via USB.

For satellite links, we deploy a Starlink RV satellite terminal. We choose this terminal because it can be moved to different locations. The terminal weighs 4.2 Kilograms and cannot be carried by our under-canopy robots or drones, highlighting the need for CBRS as the on-farm link.

\para{Coverage Area:} First, we ascertain the maximum range of the CBRS network. We move a CBRS dongle away from the base station in several directions and record the locations where the signal gets lost. All experiments for coverage are done above the crop canopy to ascertain the maximum range of the system. This allows us to determine the boundaries of the network. The results are shown in Fig.~\ref{fig:coverage}. We find that clients can successfully connect to the base station at a distance of at most 3.6 Km away. Note that the area corresponds to over 2000 acres. Therefore, without crops, a CBRS base station can cover large parts of the farm without movement. %

\begin{figure*}[!t] %
    \centering
    \subfloat[]{%
        \includegraphics[width=0.245\textwidth]{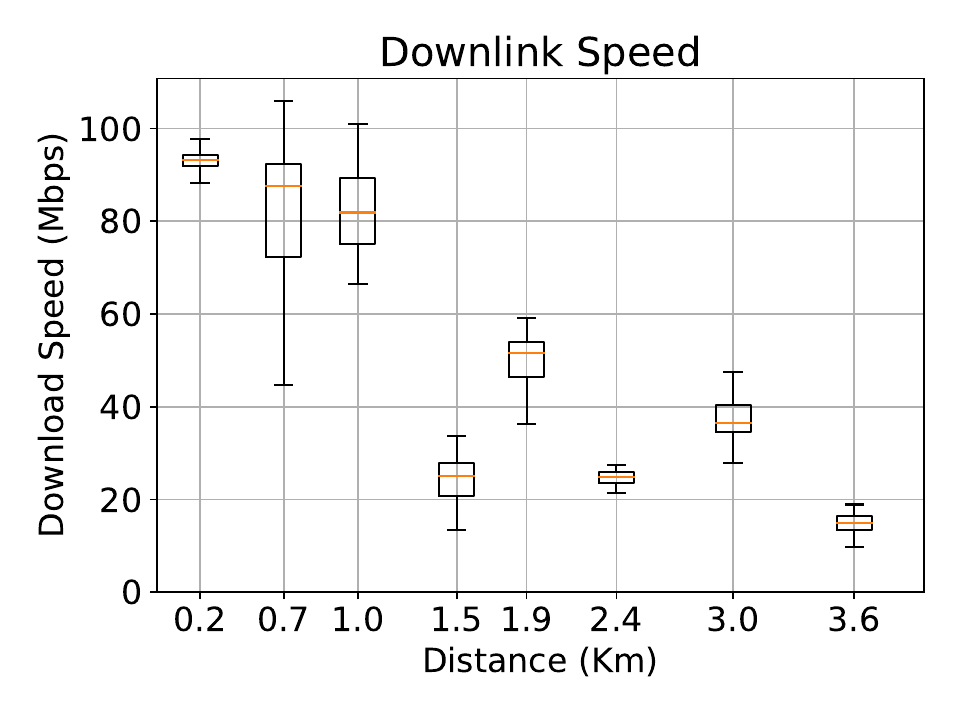} %
        \vspace{-0.25in}
    }%
    \subfloat[]{%
        \includegraphics[width=0.245\textwidth]{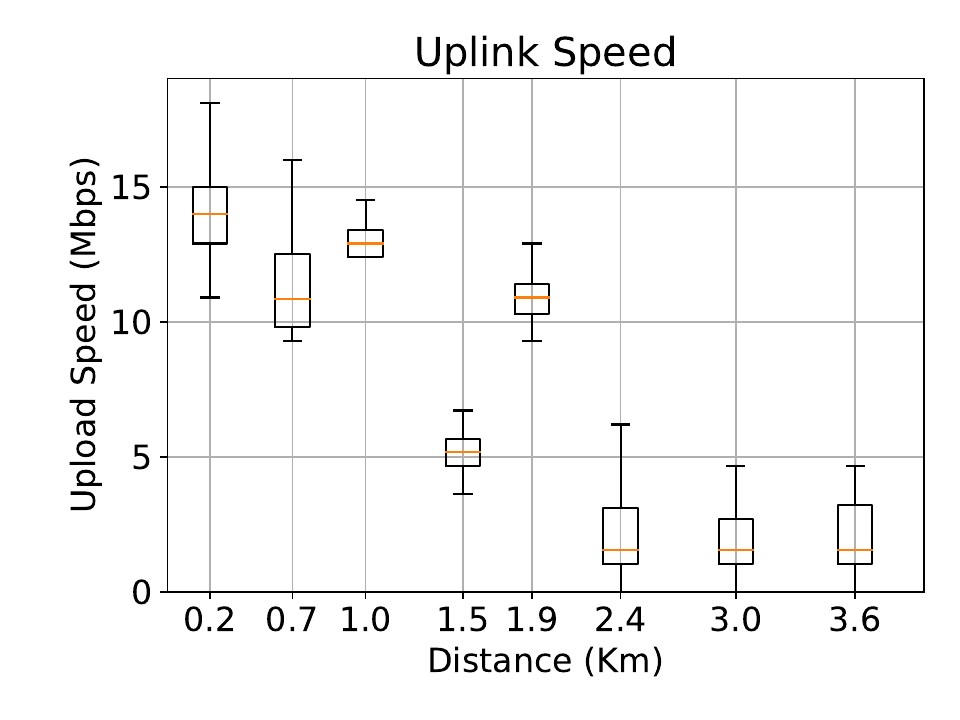} %
       \vspace{-0.25in}
    }%
    \subfloat[]{%
        \includegraphics[width=0.245\textwidth]{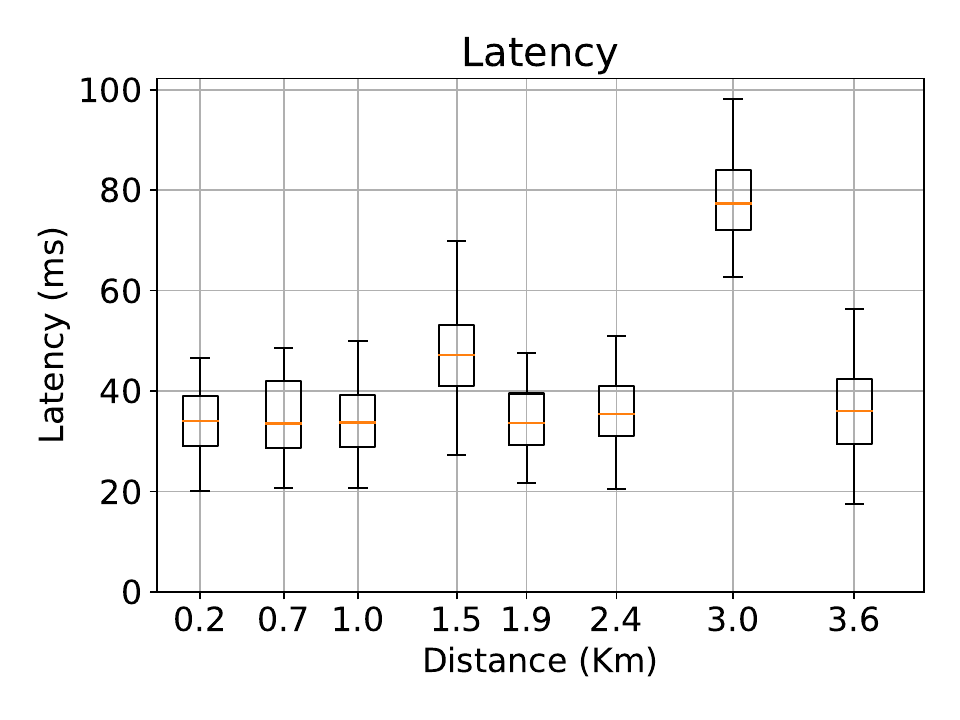} %
        \vspace{-0.25in}
    }%
    \subfloat[]{%
        \includegraphics[width=0.245\textwidth]{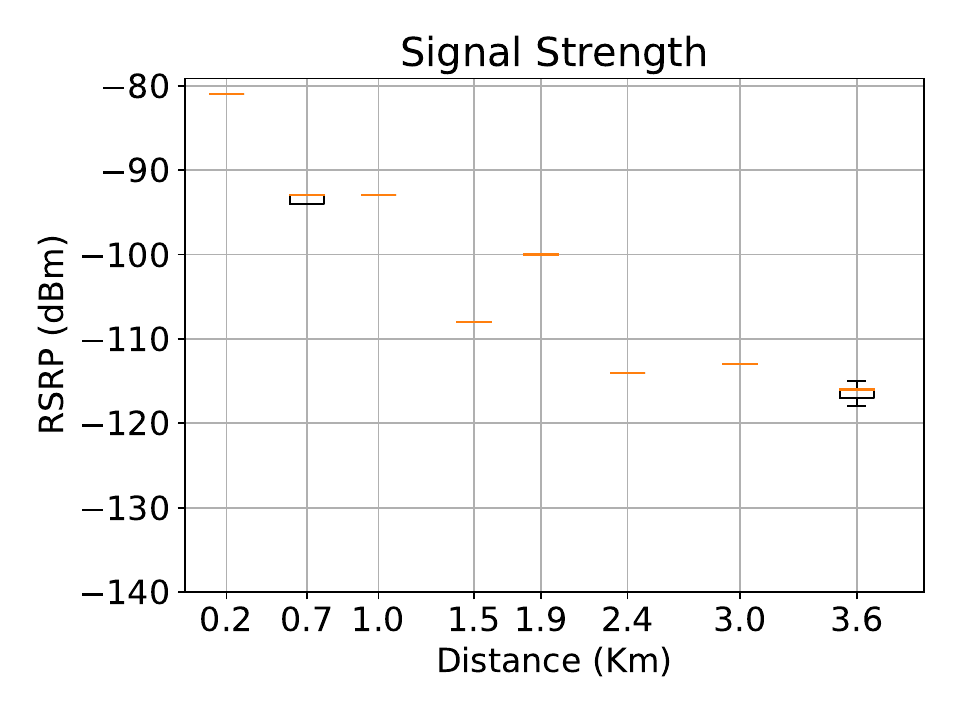} %
        \vspace{-0.25in}
    }%
    \vspace{-0.15in}
\caption{Profiling the relationship between downlink, uplink, latency, and RSRP versus distance \textbf{without} crops.}

\label{fig:vs_distance}
    \vspace{-0.25in}
\end{figure*}

\para{Connectivity Without Crops:} Next, we sample eight locations at varying distances within the coverage area that have a direct line of sight with the base station. In these experiments, the client device is placed above the crop canopy to avoid any crop-related losses. At these locations, we measure the throughput, latency, and reference signal received power (RSRP) reported by the dongle. The results are shown in Fig.~\ref{fig:vs_distance}. The signal strength degrades with distance as expected. The overall variation is around 40 dB across 3.6 Km. The downlink and uplink speeds degrade accordingly. We observe irregular patterns at a subset of the locations. We suspect that this is due to multipath reflections. Finally, note that the downlink speeds are higher than uplink speeds although they follow similar trends. This is largely due to power asymmetry between the base station and clients i.e. the base station transmits much higher power than the clients.

\begin{figure*}[!t] %
    \centering
    \subfloat[]{%
        \includegraphics[width=0.24\textwidth]{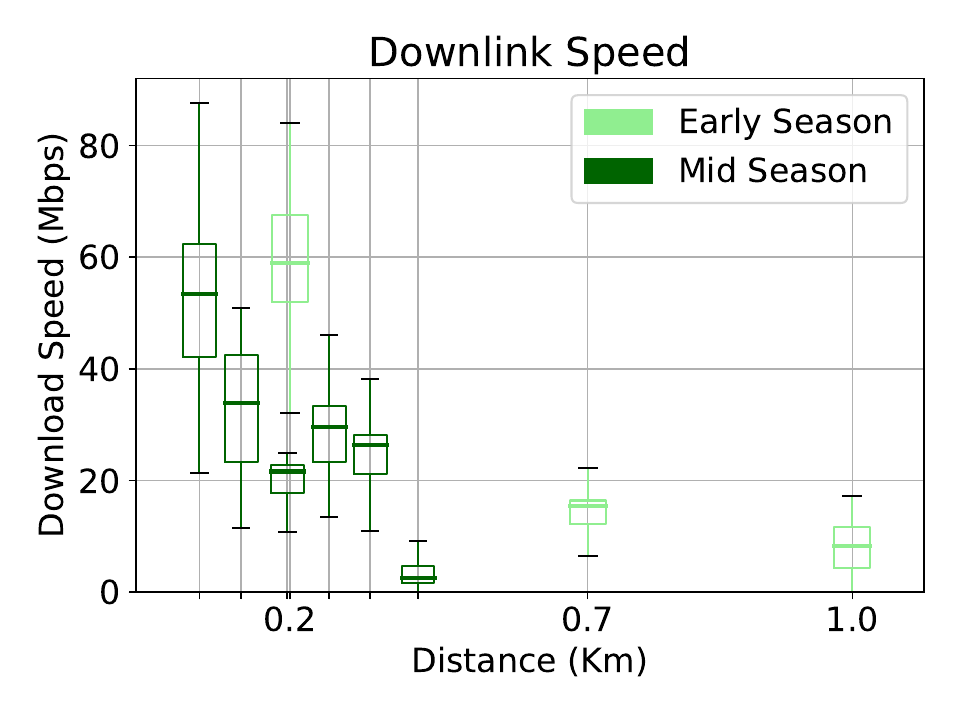} %
        \vspace{-0.15in}
    }%
    \hfill
    \subfloat[]{%
        \includegraphics[width=0.24\textwidth]{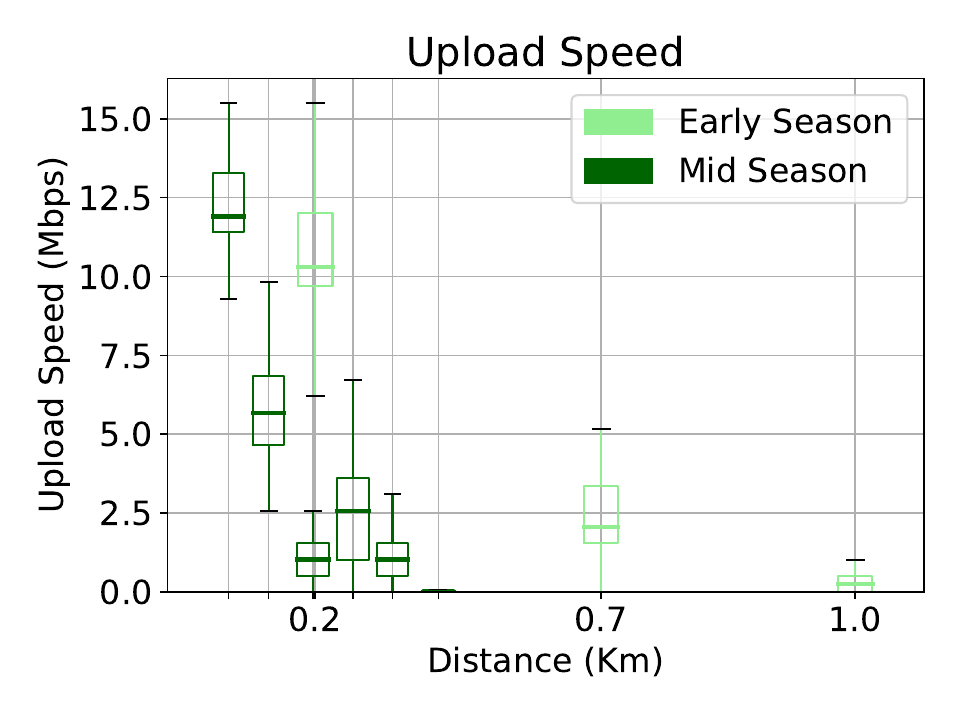} %
        \vspace{-0.15in}
    }%
    \hfill
    \subfloat[]{%
        \includegraphics[width=0.24\textwidth]{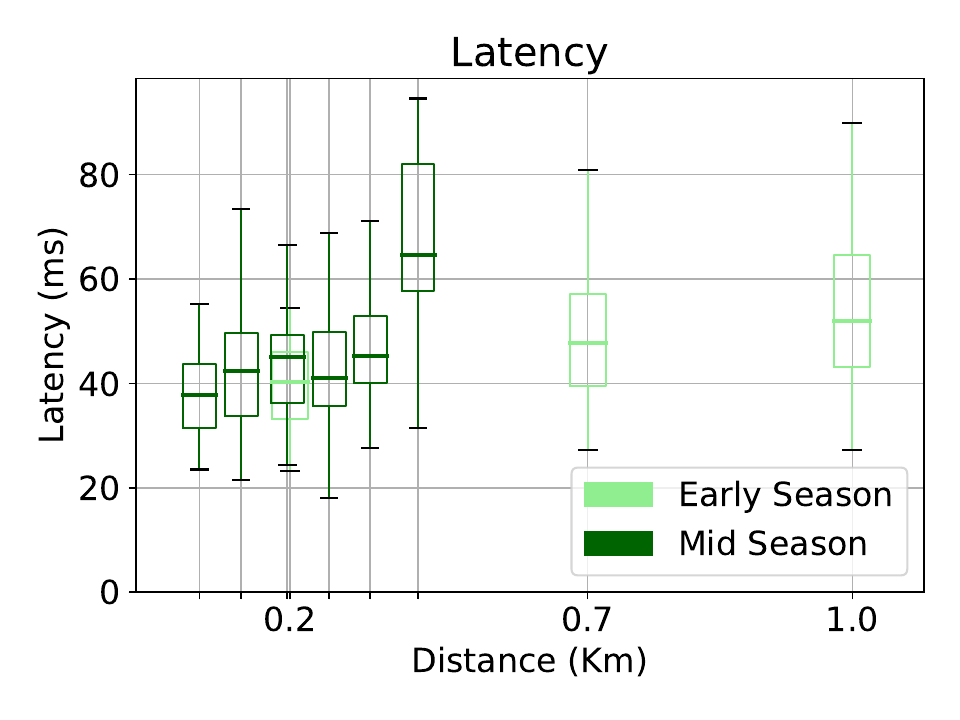} %
        \vspace{-0.15in}
    }%
    \hfill
    \subfloat[]{%
        \includegraphics[width=0.24\textwidth]{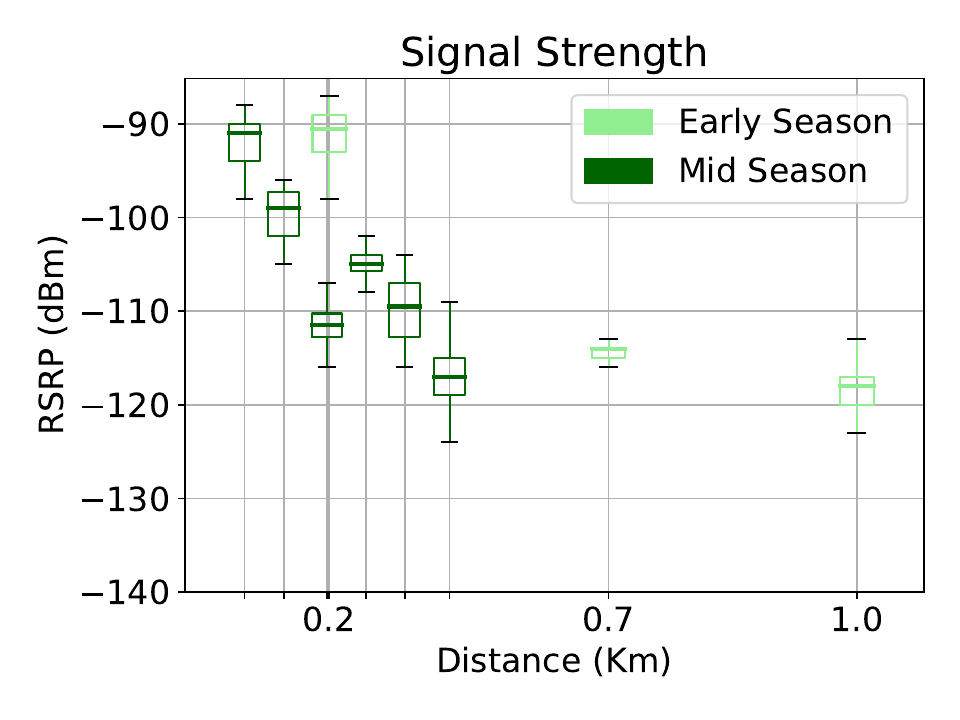} %
        \vspace{-0.15in}
    }%
\vspace{-0.15in}
\caption{Profiling the relationship between downlink, uplink, latency, and RSRP versus distance \textbf{through} crops.}
\vspace{-0.2in}
\label{fig:vs_distance_crops}
\end{figure*}

\begin{figure*}[!t] %
    \centering
    \subfloat[]{%
        \includegraphics[width=0.32\textwidth]{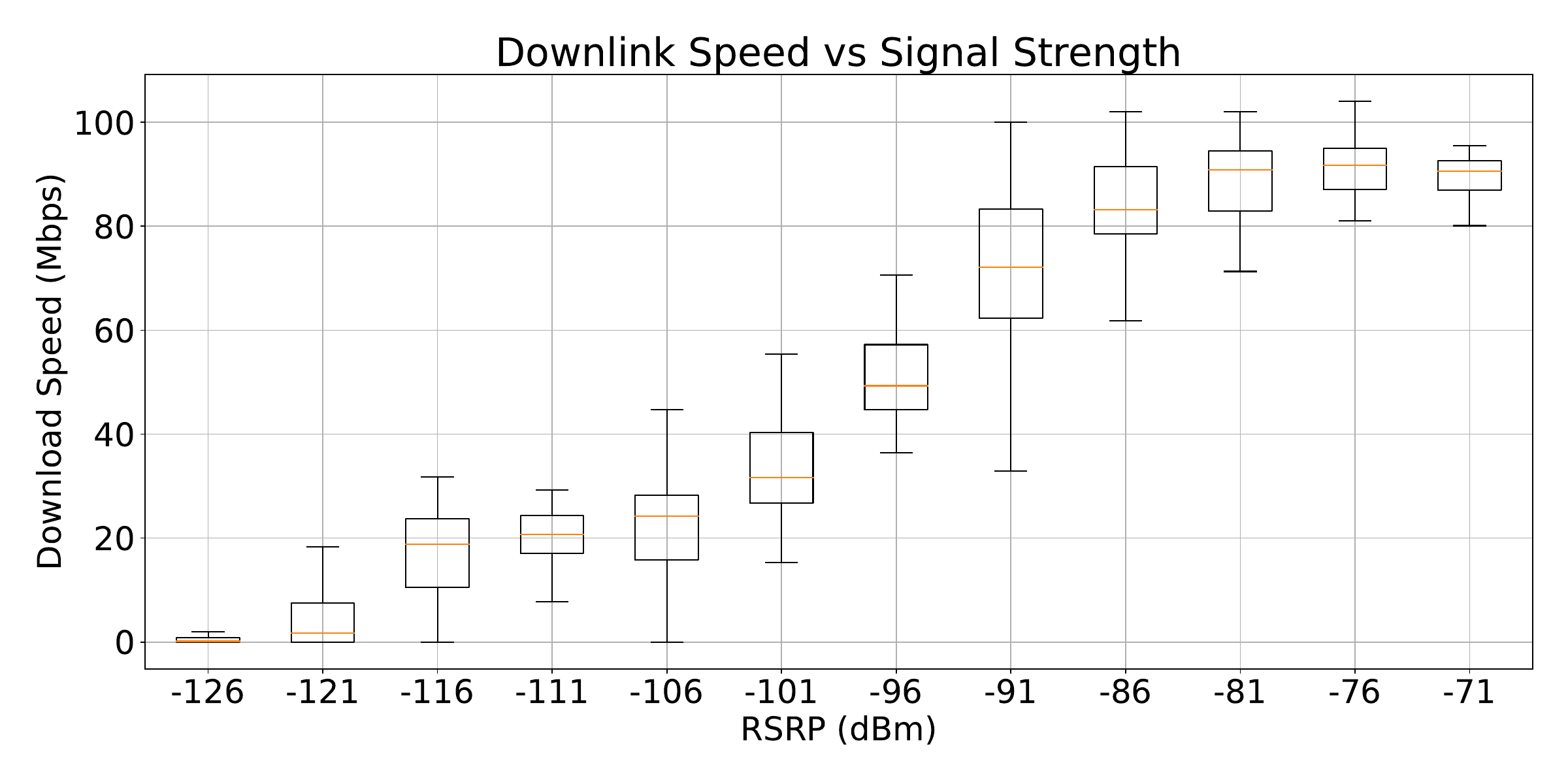} %
        \vspace{-0.15in}
    }%
    \hfill
    \subfloat[]{%
        \includegraphics[width=0.32\textwidth]{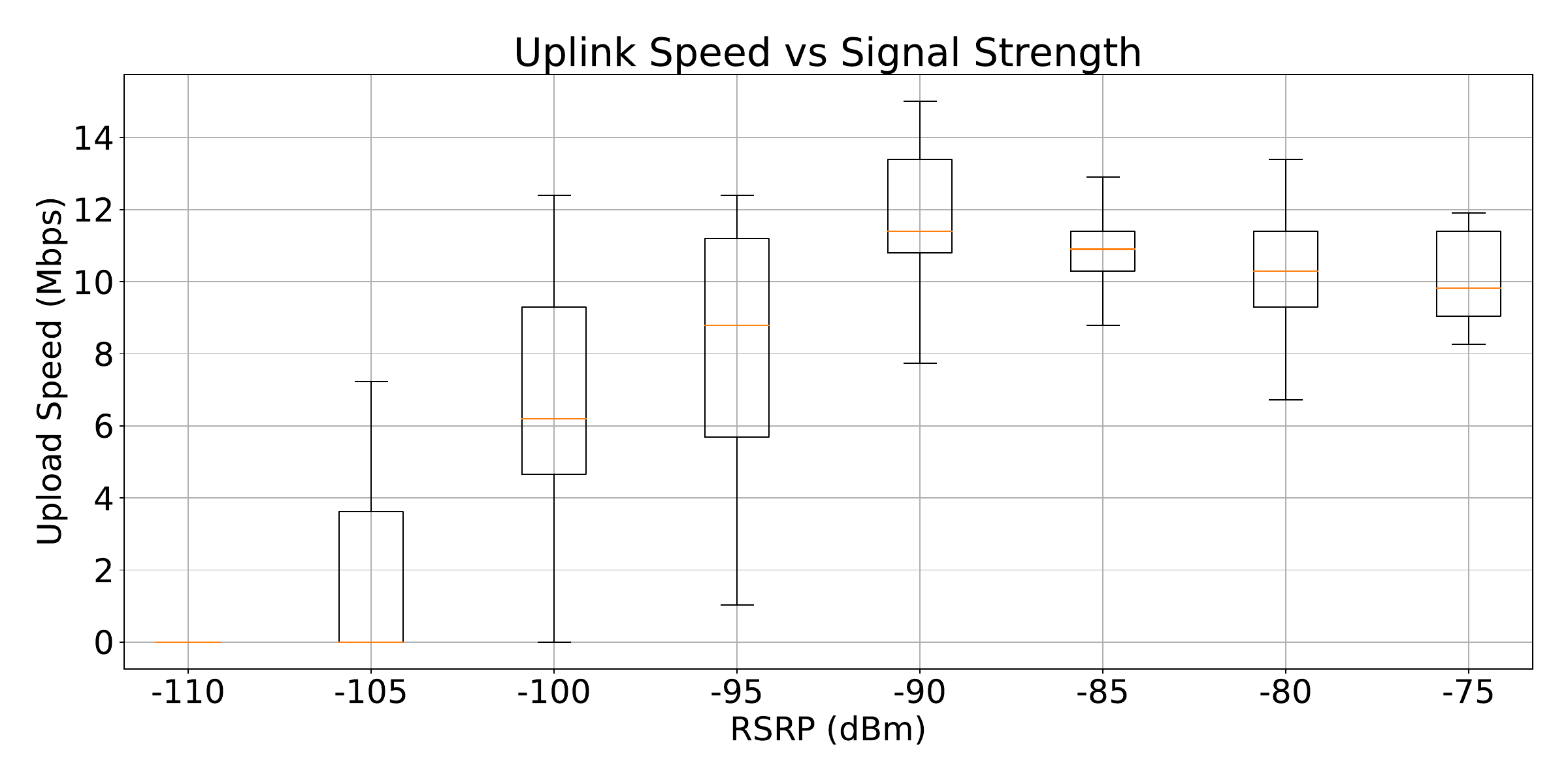} %
        \vspace{-0.15in}
    }%
    \hfill
    \subfloat[]{%
        \includegraphics[width=0.32\textwidth]{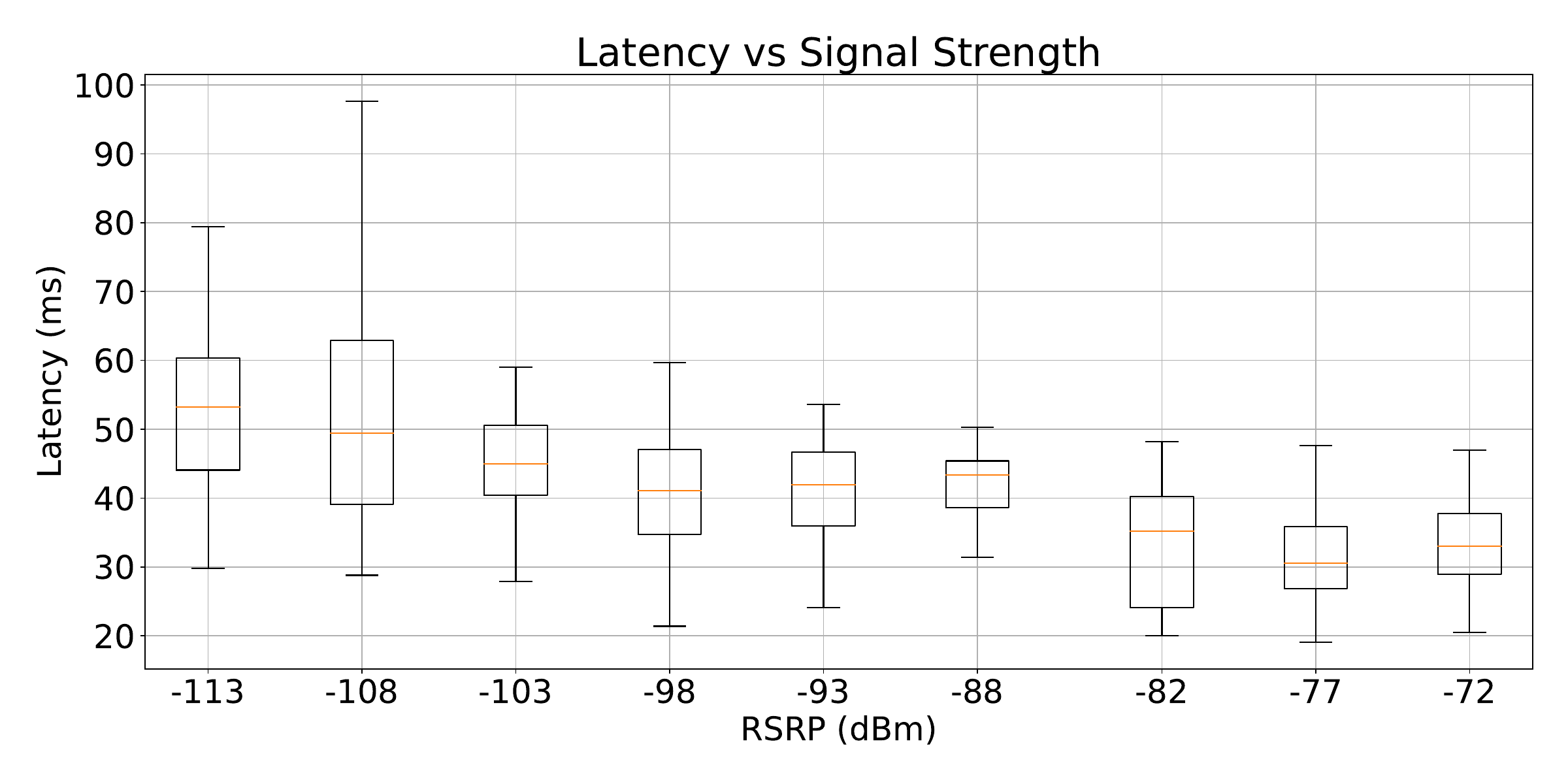} %
        \vspace{-0.15in}
    }%

\vspace{-0.15in}
\caption{Profiling the relationship between the downlink, uplink, and latency versus RSRP.}
\label{fig:vs_signal}
\vspace{-0.15in}
\end{figure*}

\para{Connectivity Through Crops:} We repeat the previous experiment with under-canopy client devices. We perform this experiment twice in the season, once with light crop cover and another later in the season with dense crop cover and plot this in Fig.~\ref{fig:vs_distance_crops}. We make three observations:
\squishlist
\item\textbf{Crops Degrade Throughput: } Under-canopy throughput, both uplink and downlink, is significantly lower than over-canopy throughput. This is because crops degrade CBRS signals. Free-space attenuation depends on the distance, $d$, between the sender and receiver and varies as $\frac{1}{d^2}$. Crops cause exponential attenuation due to their electrical permittivity and lead to the degraded throughput.

\item\textbf{Reduced Coverage: }As a corollary of the above, under-canopy coverage is reduced to around 0.2 Km and 0.75 Km in dense and light crop cover respectively. Despite the low coverage, it should be sufficient to only require infrequent movement of BYON. For example, a 0.2 Km coverage range corresponds to nearly ten acres and 0.75Km coverage corresponds to  nearly 130 acres. As we discuss in Sec.~\ref{sec:approach}, these measurements imply that a single CBRS base station is insufficient to cover a large farm. However, a BYON setup can be moved to different parts of the farm every few hours as the farming activity shifts.

\item\textbf{Seasonal Variation: }As the crops get denser, the signal obstruction due to them increases. We also expect the crop-induced variation to change with rainfall, irrigation, etc. due to variation in moisture content. 
\squishend

\para{Connectivity and RSRP:} Finally, we study the relationship between the RSRP values reported by our cellular dongle and the overall connectivity. We aggregate data points by randomly sampling points within the coverage area and plot the relationship of throughput and latency vs RSRP. The results are shown in Fig.~\ref{fig:vs_signal}. Overall, RSRP values are a good predictor of the overall connection quality. Downlink and uplink throughputs are positively correlated with RSRP and have diminishing returns as RSRP gets lower. Note that RSRP is a better predictor of downlink throughput than uplink throughput as RSRP is only measured by the client on the downlink channel. Both latency and jitter tends to decrease with increasing RSRP.

\para{Satellite Link: }We measure the uplink and downlink throughput for our Starlink terminal in outdoor setting with clear access to the sky. We measure an uplink throughput up to 50-60 Mbps, and a downlink throughput up to 200 Mbps, with temporal variations caused by satellite orbits and weather. Both uplink and downlink measurements supersede the CBRS throughput. This indicates that Starlink is a capable backhaul for the CBRS base station. In the rest of the paper, we focus on CBRS as the bottleneck link.

Finally, we place the satellite terminal in a location covered by crops. The terminal cannot connect to the satellite because satellite-ground link use high frequencies (over 10 GHz) which are easily blocked by crops. This further validates the intuition that we cannot directly equip under-canopy devices like robots with satellite links. We need a CBRS base station to provide on-farm connectivity.

\section{Benefits of Horizontal Motion}
\label{sec:motion}

\begin{figure*}[t!] %
    \centering
    \subfloat[CBRS network coverage with one sector antenna. White: without crops. Green: with crops.]{%
        \includegraphics[width=0.28\textwidth]{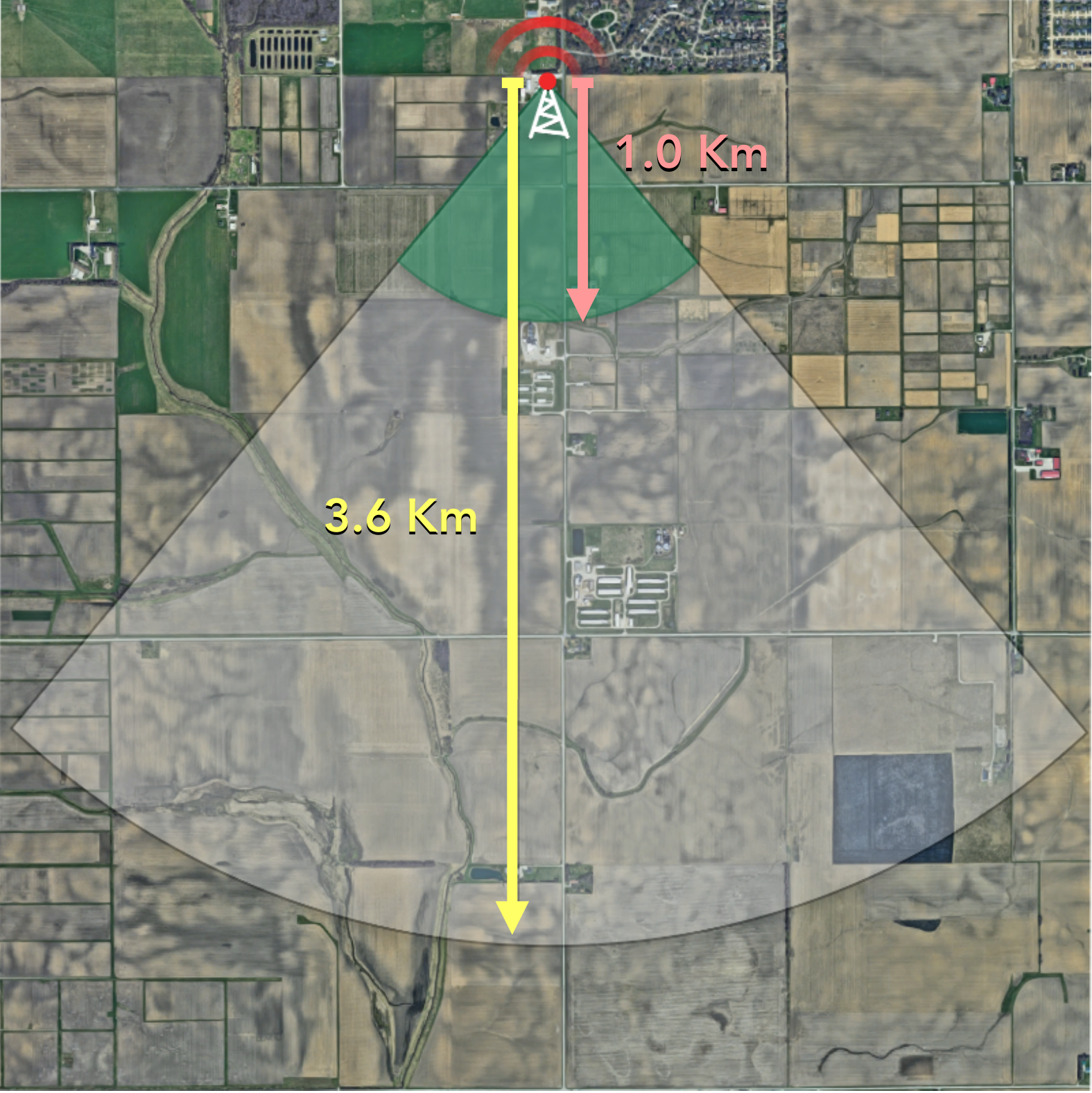}
        \label{fig:coverage_area}
    }%
    \hfill
    \subfloat[Covering a $4.8 \times 4.8$ km farm by replicating deployment in \ref{fig:coverage_area}. Each bubble has radius 1 km.]{%
        \includegraphics[width=0.30\textwidth]{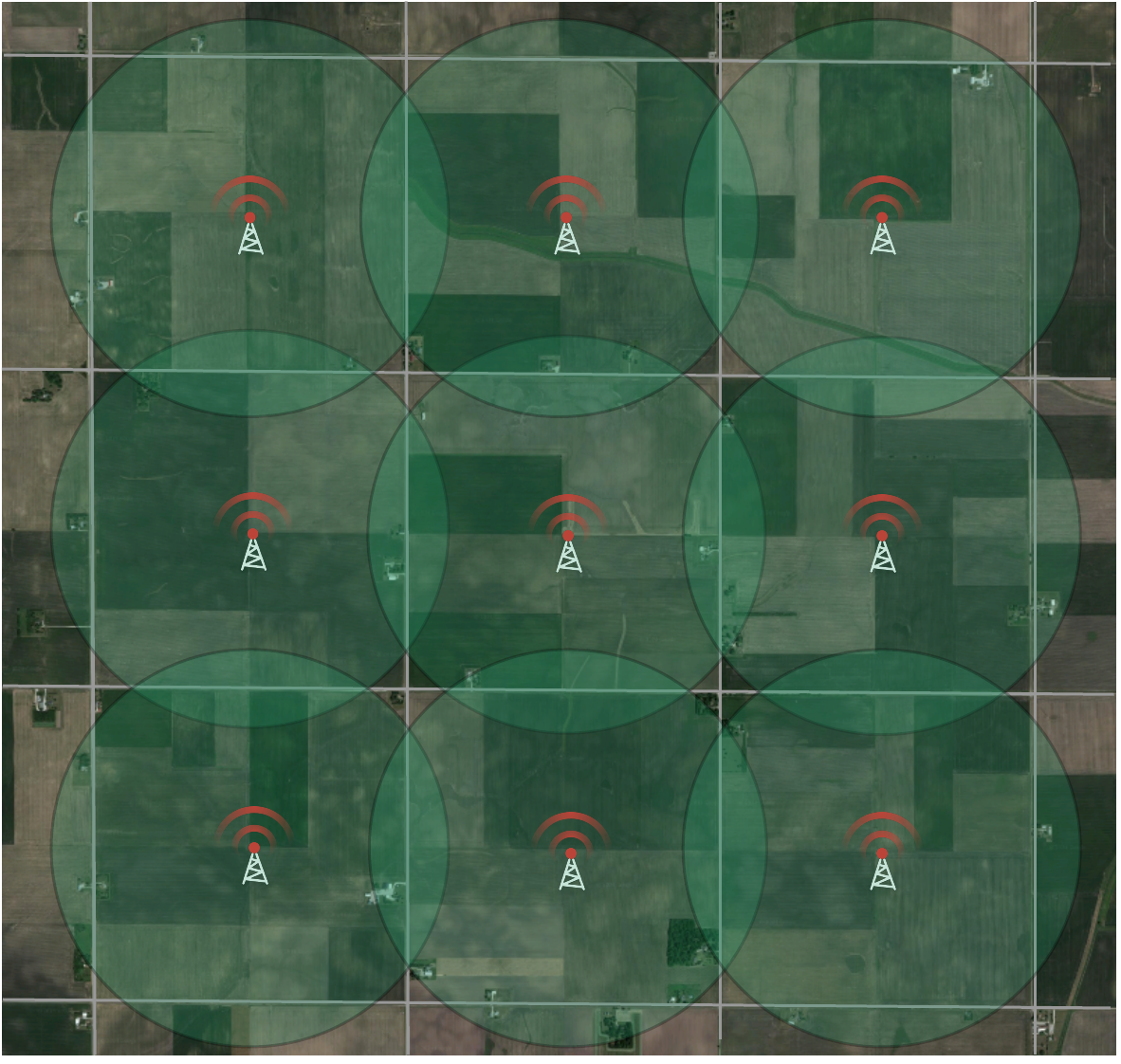} 
        \label{fig:large_farm}
    }%
    \hfill
    \subfloat[Covering the same farm w/ \name. Any point can be covered by moving \name\ appropriately.] {%
        \includegraphics[width=0.30\textwidth]{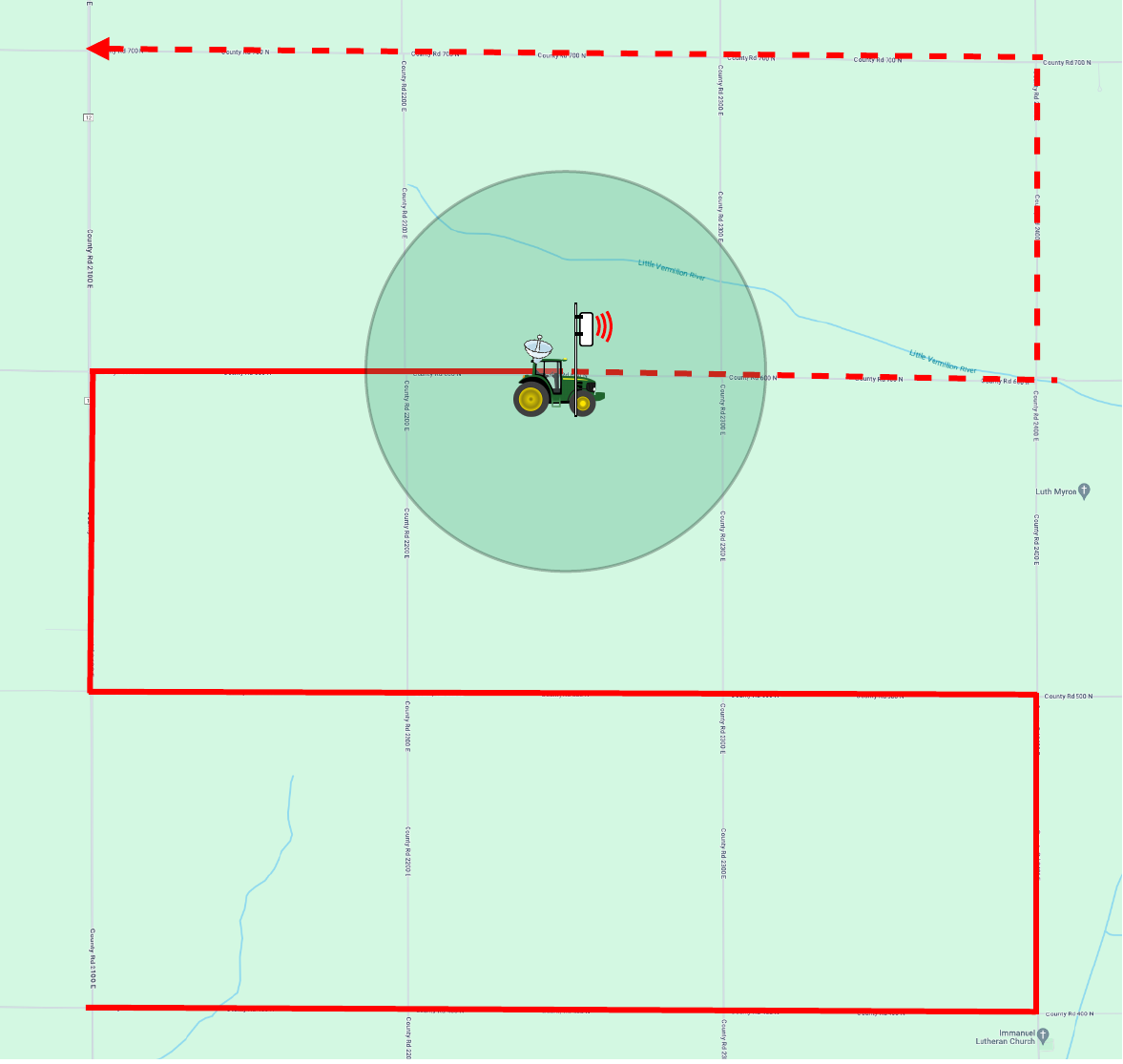}
        \label{fig:horizontal_motion}
    }%
\vspace{-0.15in}
\caption{Extending wireless coverage across a large-scale farmland. By observing that digital agriculture applications are localized in space and time, BYON can achieve large decreases in infrastructure cost when compared with conventional static infrastructure. }\vspace{-0.15in}
\label{fig:coverage}
\end{figure*}
\noindent\textbf{{Seasonal Rhythms of Agriculture:}} Our key observation is that the connectivity requirements in agriculture are restricted in space and time. For example,~\cite{iowaextension} lists some activities that use equipment and how long they take. In an 800 acre field planting corn, pre-planting activity such as nitrogen application would take about 8 days with the tractor covering about 96 acres per day (0.4 $Km^2$). Similarly, planting takes about 3 days, covering approx. 280 acres per day. At harvest time, harvesting 800 acres takes about 11 days, i.e., 70 acres per day. The average farm size in the United States is 441 acres~\cite{agcensus}. Similarly, a farm robot covers about 20-50 acres per day~\cite{laserweeds,airobotweeds}. Therefore,  for an average farm, farm activity, especially those involving equipment is limited to small parts of the farm on a given day. Moreover, this part of the farm shifts across time in a season.

\para{Analysis: }We quantify the cost benefits of horizontal motion by considering a large Midwestern farm area in United States as shown in Fig.~\ref{fig:large_farm}. This farm area measures $4.8 \times 4.8$ km, which would correspond to a large 5000-acre farm in United states. According to our measurement study, the maximum extent of coverage through crops during the peak season is roughly 1 km as shown in Fig.~\ref{fig:coverage_area} given a fixed antenna height of 5m (Fig.~\ref{fig:deployment}). Hence, to cover this farm throughout the season by replicating this setup, we would require doing so in a $3 \times 3$ grid-like arrangement as shown in Fig.~\ref{fig:large_farm}. As mentioned previously, this solution is over-provisioned as only a small number of fields within the farm will need high-bandwidth connectivity at any given moment. On the other hand, by horizontally moving \name, we can selectively choose to extend coverage to different areas of the farm as needed  (Fig.~\ref{fig:horizontal_motion}). Hence, by leveraging horizontally moving base stations, we can achieve drastic cost savings when compared to conventional static infrastructure.

Although horizontal motion allows us to move bubbles of connectivity, it does not allow us to optimize the connectivity within the bubble (i.e. deal with crop-induced attenuation). To achieve this, \name\ leverages vertical motion in addition to horizontal motion. We describe details of our variable-height base station in the following section (Sec.~\ref{sec:design}).

\section{Variable-Height Base Station}
\label{sec:design}

\begin{figure}[!t]
    \centering
    \includegraphics[width=.9\linewidth]{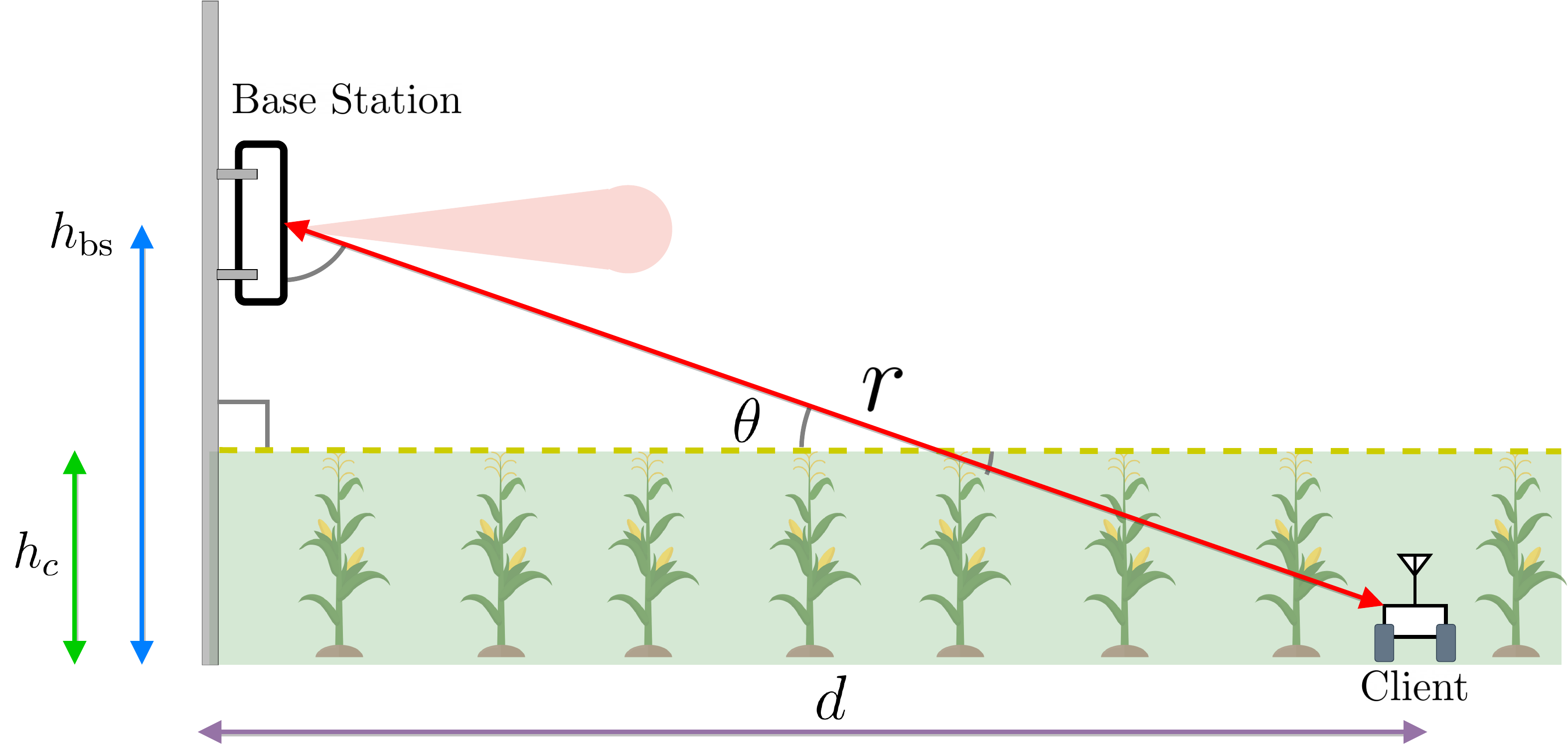}
    \vspace{-0.15in}
    \caption{Client under crop canopy of height $h_c$ at horizontal distance $d$ from base station at height $h_{bs}$.}
    \label{fig:scenario}
    \vspace{-0.15in}
\end{figure}

In this section, we aim to optimize throughput for under-canopy applications by designing a variable height base station. Note this does not incur additional hardware cost: standard vehicle-mounted antenna masts (e.g. on CoWs) are already telescoping as they need to be stowed during transit or storage and quickly retractable during high wind conditions \cite{aluma_smarttower}. We discuss the intuition and the formulation for the variable-height base station below.

\subsection{Exploiting Height Variability} 
First, we explain the advantage of dynamic height variability. Consider the basic scenario in Fig.~\ref{fig:scenario}. Suppose there are no crops i.e. $h_c=0$. Then the optimal base station height should be close to the ground i.e. $h_{bs} \approx 0$ as it minimizes the physical distance $r$ between the base station and the client. Specifically, 
\begin{equation}
r = \sqrt{d^2 + h_{bs}^2}
\end{equation}

On the other hand, suppose there are crops i.e. $h_c > 0$. If we set $h_{bs} \approx 0$, i.e., very close to the ground, the direct path to the client will be horizontal and cover a large distance $r\approx d$ through the crops. This will lead to large attenuation for the signal and is sub-optimal.

To counter this problem, one may be inclined to increase $h_{bs}$. When $h_{bs}$ is very large, the distance travelled through crops gets increasingly closer to $h_c$, which is the minimum distance the signal can cover through crops. Therefore, increasing the height of the base station has the advantage of reducing through-crop attenuation.

However, as we increase $h_{bs}$, two other effects occur. First, the total propagation distance $r$ increases. Second, the angular deviation from the base station antenna's main lobe to the client, denoted by $\theta$, increases. $\theta$ is related to $h_{bs}$ and $d$  by \vspace{-0.20in}
\begin{equation}
\theta = \arctan\biggl(\frac{h_{bs}}{d}\biggr)
\end{equation}%
Both factors negatively affect received power at the client. In other words, there is tradeoff regarding $h_{bs}$. If it is too low, the connection will suffer from severe crop blockage. If it is too high, the connection will suffer from path loss and antenna-related losses i.e. $r$ and $\theta$ are too large. Hence, in the presence of crops, there should be an optimal base station height which balances these two competing factors.

\para{Why Not Raise the Client Antenna?} It is tempting to modify the under-canopy client to have antennas extending above the crop canopy. While the robot can travel through the space at the crop base, the canopy is dense as shown in Fig.~\ref{fig:crop-canopy}. Therefore, an antenna extending above the crop canopy will drive/cut through the crops, causing damage to the crops and contributing large drag forces onto the robot.

\begin{figure*}[!t] %
  \centering
  \subfloat[Schematic. Note the correspondence with Fig.~\ref{fig:scenario}.]{%
    \includegraphics[width=0.4\textwidth]{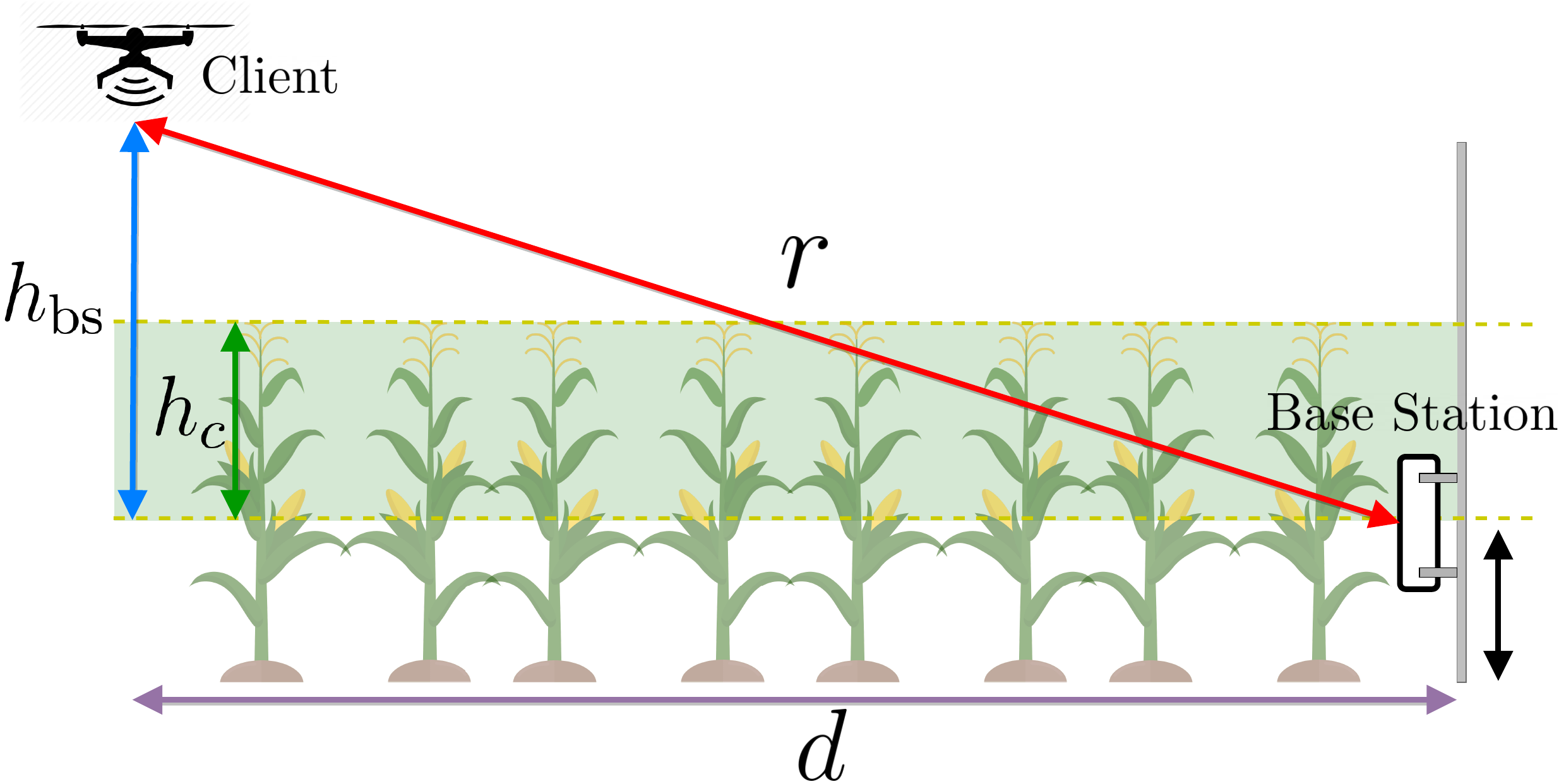} %
    \vspace{-0.3in}
  }%
  \hfill
  \subfloat[Antenna post used to vary $h_c$.]{%
    \includegraphics[width=0.28\textwidth]{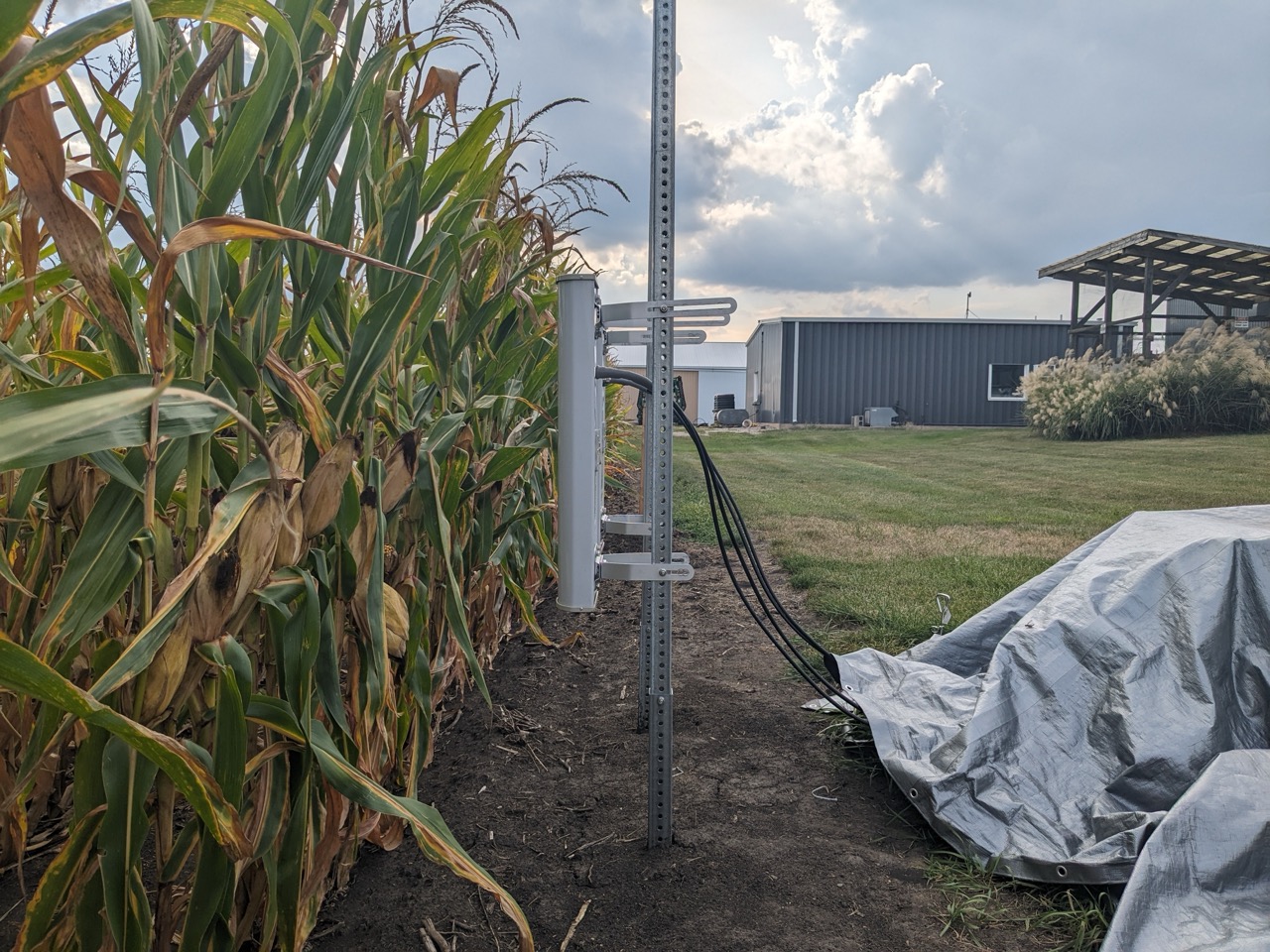} %
    \vspace{-0.3in}
  }%
  \hfill
  \subfloat[UAV with CBRS dongle used to vary $h_{\text{bs}}$.]{%
    \includegraphics[width=0.28\textwidth]{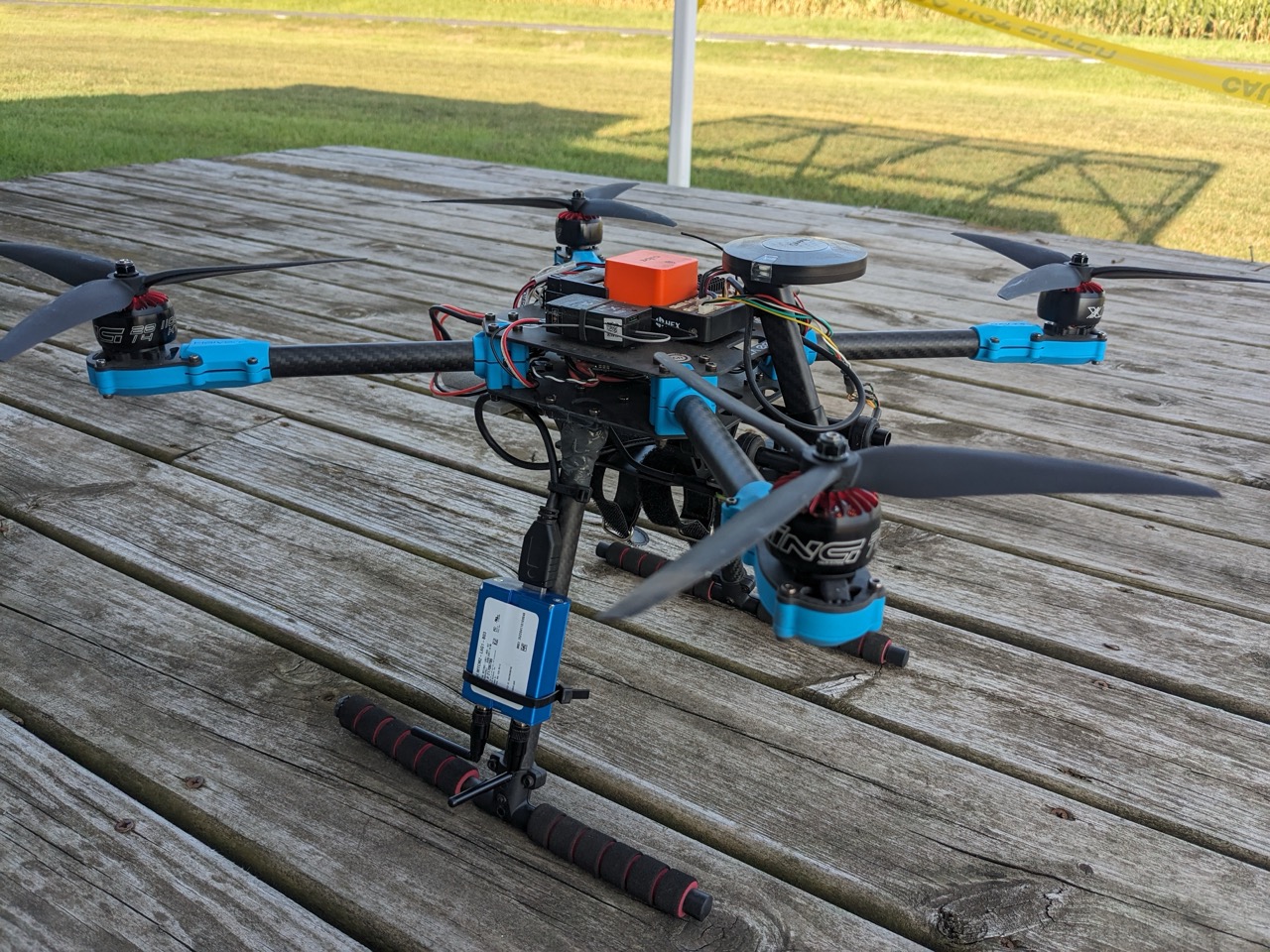} %
    \vspace{-0.3in}
  }
  \vspace{-0.1in}
  \caption{\textbf{Experimental Setup.} We swap the positions of the base station and client using channel reciprocity. Our client mounted on a UAV acts as a surrogate base station, while our base station acts as a surrogate client.}
  \label{fig:experimental_setup}
\vspace{-0.25in}
\end{figure*}

\begin{figure*}[!t] %
  \centering
  \subfloat[Without crops (i.e. $h_c=0$), the signal strength monotonically decreases with height.]{%
    \includegraphics[width=0.32\textwidth]{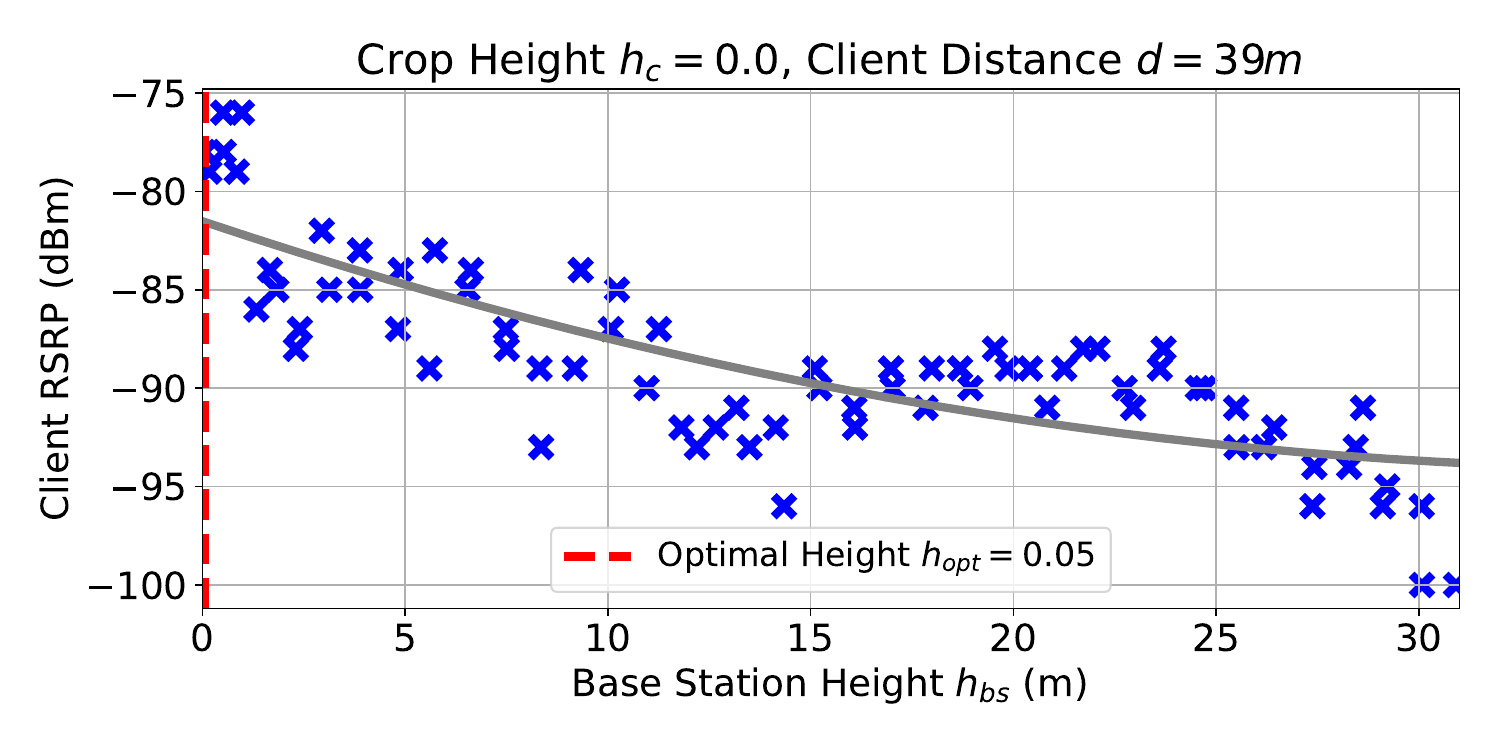} %
    \vspace{-0.3in}
  }%
  \hfill
  \subfloat[With crops (e.g. $h_c=1.0$m), the signal strength increases to an optimum, then decreases.]{%
    \includegraphics[width=0.32\textwidth]{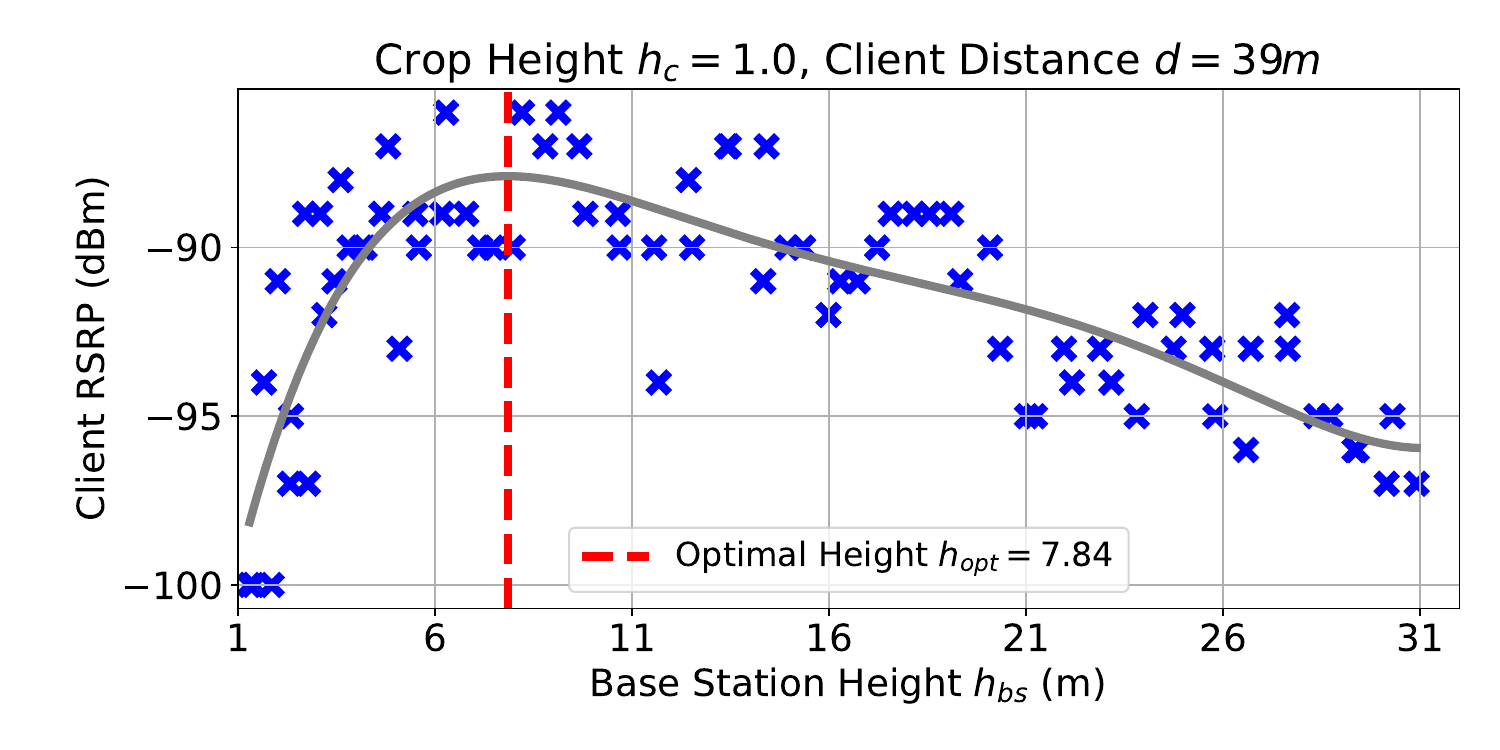} %
    \vspace{-0.3in}
  } 
  \hfill
  \subfloat[Optimal height increases with client distance $d$.]{%
    \includegraphics[width=0.32\textwidth]{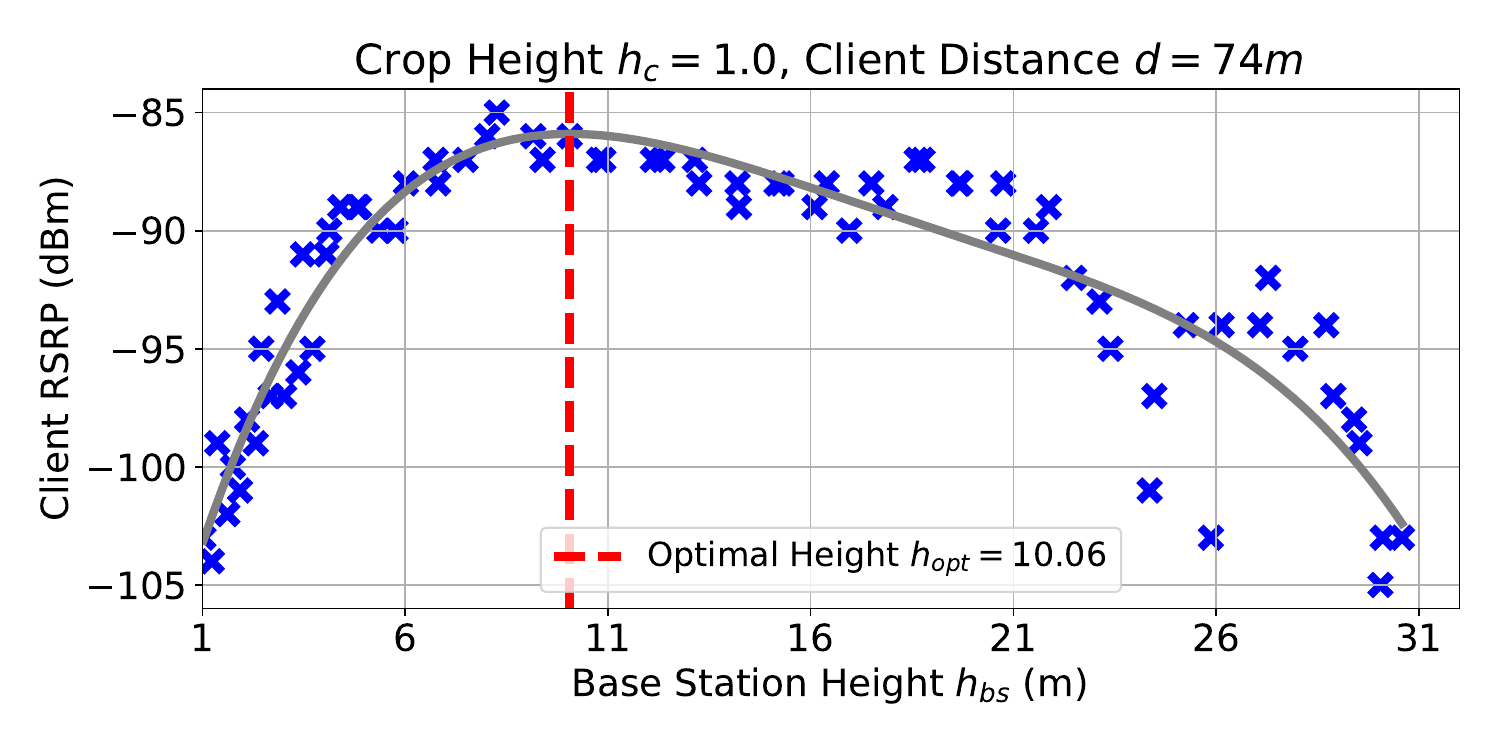} %
    \vspace{-0.3in}
  }
  \vspace{-0.1in}
  \caption{\textbf{Intuition for the optimal height.} We place a client at a horizontal distance $d$ from the base station and vary the height of the base station $h_{\text{bs}}$ by flying the drone vertically up and down (see Fig.~\ref{fig:experimental_setup}). We scatter plot the observed RSRP at the client vs the base station height. The best-fit line is shown in grey. The optimal heights are marked as dotted red lines.}
  \label{fig:intuition}
\vspace{-0.2in}
\end{figure*}

\subsection{Experimental Validation}

To validate these ideas, we perform an experiment using the setup in Sec.~\ref{sec:eval_setup} and  Fig.~\ref{fig:experimental_setup}. We emulate a fixed client at $d=39$ m and vary the value of $h_{bs}$ for different crop heights, $h_c$. We measure the RSRP values on the client and plot the results in Fig.~\ref{fig:intuition}(a)(b). The experiment supports our intuition: (a) The RSRP degrades continuously with distance in the absence of crops, i.e., the optimal $h_{bs} \approx 0$ without crops. (b) In the presence of crops, the signal is weak when $h_{bs}$ is low and gets stronger with increasing heights. However, beyond an optimal value of $h_{bs} > 0$, the signal gets weaker again. Note that changing the height of the base station can significantly impact the RSRP of the received signal (up to 15 dBm variation as shown in Fig.~\ref{fig:experimental_setup}(b)). In Fig.~\ref{fig:vs_signal}, we show how such RSRP variations map to throughput variations. %

Finally, we consider what happens when varying the client distance $d$. Note that with increasingly large $d$, the influence of $h_{bs}$ on both $r$ and $\theta$ becomes increasingly marginal. Hence, we are incentivized to raise the base station height further as this does not incur much of an increase on $r$ and $\theta$. To summarize, increasing client distances favor increasing base station heights. This is validated by our experiment in Fig.~\ref{fig:intuition}(c). As our clients are mobile devices, this further substantiates the need for a dynamic height-varying base station.

\subsection{Modeling the Dynamic Base Station} 

Given the performance variation due to height of the base station, we need to define a mechanism to identify the optimal height of the base station. One option is to have the base station probe different heights. In practice, such a system is bound to be slow in reacting to the robot motion under crops as it involves mechanical motion. Our approach is to derive an explicit physics-based model that predicts the client RSRP given a base station configuration. Once we have this model, we can use it to directly predict the optimal height of the base station, given crop height.

Our model incorporates 3 major factors that account for the variation of the CBRS signal strength:

\para{(i) Path Loss:} We know that the signal becomes weaker when the receiver is further away from the base station. In vacuum and air, the energy of the signal $E$ and the distance between the sender and receiver $r$ generally follows $E\propto r^{-2}$~\cite{electrodynamics_textbook}. Therefore, the path loss is%
\begin{equation}
    \label{eqn:pass-loss}
    L_P=-20\log_{10}r%
\end{equation}
\para{(ii) Crop Attenuation:} When the wireless signal travels through crops, it suffers additional attenuation because part of the signal is deflected and absorbed by the crops. When the signal is not excessively strong, the absorption can be regarded as linear, i.e. the energy absorption is proportional to the current energy%
\begin{equation}
    \label{eqn:energy-absorption}
    \frac{dE}{dr}=-\alpha E%
\end{equation}
The energy will therefore experience an exponential decay, i.e. $E\propto e^{-\alpha r_c}$ where $r_c$ is the distance in crops. Therefore, the loss due to crop absorption is%
\begin{equation}
    \label{eqn:absorption-loss}
    L_{A}=-\alpha r_c%
\end{equation}
$\alpha$ is an absorption coefficient that depends on the crop type, and density. Using trigonometry, we have: $r_c = \frac{h_c}{h_{bs}} r$. 

\para{(iii) Antenna Directivity:} The antenna of the base station are designed to be directional, i.e. the radiation energy does not distribute evenly on all angles. In fact, the CBRS base station we are using in the evaluation has a radiation pattern similar to the one depicted in Figure~\ref{fig:scenario}. It has a relatively flat radiation pattern in the azimuthal dimension, but a constrained radiation pattern in the elevation dimension. Therefore, the signal strength will vary based on the relative angle of elevation ($\theta$) between the base station and the device. Generally, base stations use an antenna array to perform beam forming~\cite{shepard2012argos} to concentrate its energy on a narrow beam. This would yield a beam pattern of the $\sinc$ function. We thus model the loss~(gain) due to angle as%
\begin{equation}
    \label{eqn:angle-loss}
    L_{D}=10\log_{10}\left(\max\left(|\sinc\left(\Gamma\theta\right)|,\beta\right)\right)%
\end{equation}
which is a $\sinc$ function capped at a minimum value $\log\beta$ for numerical reasons.

Combining the factors, the total signal strength~(RSRP) is%
\begin{equation}
    P_{\alpha, \beta, \Gamma, G}(r, \theta, r_c) = L_{\alpha, \beta, \Gamma}(r, \theta, r_c) + G%
\end{equation}
where $G$ is the lumped gain~(constant) that encapsulates all the gains in system (i.e. from the amplifier and the antenna). Moreover, $L_{\alpha, \beta, \Gamma}(r, \theta, r_c)$ represents all the losses in the system and is defined as:%
\begin{equation}
    \label{eqn:signal-model}
    L_{\alpha, \beta, \Gamma}(r, \theta, r_c) = \underbrace{-20 \log_{10} r}_{\text{Path Loss}} \underbrace{- \alpha r_c}_{\text{Crop Attenuation}} + \underbrace{10 \log_{10} \Phi_{\beta,\Gamma}(\theta)}_{\text{Directivity}}\vspace{-0.01in}
\end{equation}
Finally, $\Phi_{\beta,\Gamma}(\theta)$ is the unit-normalized antenna radiation pattern defined as:%
\begin{equation}
    \Phi_{\beta,\Gamma}(\theta) = \max(|\sinc(\Gamma\theta)|, \beta)%
\end{equation}

Note that, our model does not capture multipath effects. This is because farm areas are open spaces and typically do not have large reflectors. We empirically validate our model in Sec.~\ref{sec:eval_model}.

\para{Estimating Model Parameters: } The power loss model defined above depends on four parameters:  $\alpha,\beta,\Gamma,G$. These are physical parameters that represent properties of the hardware and the physical environment. They can be estimated using a small number of measurements. For example, a robot can do a quick maneuver and report observed signal strength measurements along with its own positions for us to estimate these parameters. Alternatively, getting a signal from a set of already deployed sensors will provide enough data to estimate these parameters. Given a small amount of calibration data, we formulate this as a constrained non-linear least squares problem. Our objective function is to minimize the square of the error residual and our constraints are that $\alpha, \beta, \Gamma \geq 0$. We leverage trust region methods from a well-tested numerical library \cite{2020SciPy-NMeth} to perform this task. %

\para{Identifying Ideal Height of the Base Station: }We begin by considering a single target device operating in a farm, e.g., a farm robot. In this case, our model can predict the signal strength at the location of the robot for any given height. Therefore, to choose the optimal height of the base station, we simply need to identify a height that maximizes the required signal strength. To obtain the horizontal distance of the robot, the robot can send its estimated location to the base station. A robot can use GPS to self-localize whenever it is outside a crop row (i.e. when travelling between rows in Fig.~\ref{fig:phenotyping_route}). It can then continuously update its own location estimate when entering a crop row by leveraging on-board sensor readings (e.g. cameras, lidars, IMUs) \cite{sivakumar_learned_2021, velasquez_multi-sensor_2022, gasparino_cropnav_2023}.

For multiple devices with different requirements, \name\ can optimize the aggregate throughput by predicting throughput at different locations in the farm for each height. It can also optimize for other metrics of interest, e.g., throughput for a subset of devices, etc.

\section{Experimental Evaluation}
We describe our evaluation of \name's model and height-variable base station below.

\subsection{Experimental Setup}\label{sec:eval_setup}
We need to measure variation of the CBRS signal for different crop heights, different values of distance, and different heights of the base station. Moving the base station requires heavy equipment like a tractor or a truck. Such equipment can only move along edges of the field and cannot provide extensive experiments. Therefore, we perform these experiments in a flipped manner. The client is placed on a drone (Fig.~\ref{fig:experimental_setup}(C)) that dynamically varies its height and distance. The base station is placed under the crops on a set of poles (Fig.~\ref{fig:experimental_setup}(B)). The base station's height can be adjusted to observe different effective values of $h_c$. For example, to reduce $h_c$, we can simply lift the base station slightly from the ground. The schematic is shown in Fig.~\ref{fig:experimental_setup}(A). Note that, this schematic is analogous to Fig.~\ref{fig:scenario}, except that the position of the client and base station has been flipped. Given reciprocity of wireless systems, both setups yield the same measurements. The drone is equipped with an altimeter and GPS module. The altimeter and the GPS module allow us to track the drone's 3D location during flight, allowing us to measure $h_{bs}$, $d$, and by extension $r$. 

\para{Limitations of our Experimental Setup: }We note two limitations of our experiments. First, due to the lower height of the base station, we are required by regulation to reduce the power on the base station to 1W (as opposed to 50W maximum power, i.e., 17 dB lower). Therefore, as opposed to the few Km range in Sec.~\ref{sec:measurements}, our range is reduced to hundreds of meters. In practice, \name\ will be deployed using commercially available computer-controlled height-adjustable telescoping antenna masts~\cite{willburt_mast_stilleto, willburt_mast_pneumatic, aluma_smarttower} on tractors and trucks (unfortunately, we do not have access to large farm equipment since farm tractors/trucks cost upwards of 50k USD) and transmit at the full power to support larger-range operations. 
Second, we collect signal strength measurements in our experiments, not throughput because throughput measurements react slowly to the motion of the drone and do not capture instantaneous effects. However, as we see in Fig.~\ref{fig:vs_signal}, throughput and signal strength are closely related.

\subsection{Data Collection}\label{sec:dataset}

We collect an extensive dataset for our through-crop measurements using the drone setup described above. This setup allows us to emulate varying crop heights by changing the height of the base station antennas with respect to the ground, while collecting data for multiple values of $h_{bs}$ and distances. 

\para{Drone Vertical Motion Dataset: }First, we fix a crop height $h_c \in \{0.0, 0.5, 1.0, 1.5\}$m by moving the antenna vertically along the sign post. Then, we place the UAV in the field at a distance $d > 10$m away from the base station. To collect data, we fly the UAV up to a height around 30m and then back down to land. Throughout the flight, we log the GPS coordinates, altitude, and RSRP values reported by the CBRS dongle at a rate of 2 Hz. We limit the vertical speed of the UAV to 1m/s to densely sample the RSRP variation with height. After each flight, we increase $d$ by 10 to 20 meters by finding another location further away and repeat the data collection. We repeat this process until the dongle is too far away from the base station to receive a signal or we exceed the boundaries of our allotted test field. The maximum value of $d$ depends on the extent of crop blockage $h_c$. For example, with $h_c=1.5$m the maximum $d$ is around 50m. Each flight consists of roughly 1 minute of capture time and 120 data points. Our dataset consists of data from 24 such flights. After removing invalid data points due to sensor errors, our dataset comprises 2216 usable data points in total.
\begin{figure*}
    \subfloat[CDF of test error across entire test dataset.]{
        \includegraphics[width=0.35\textwidth,valign=c]{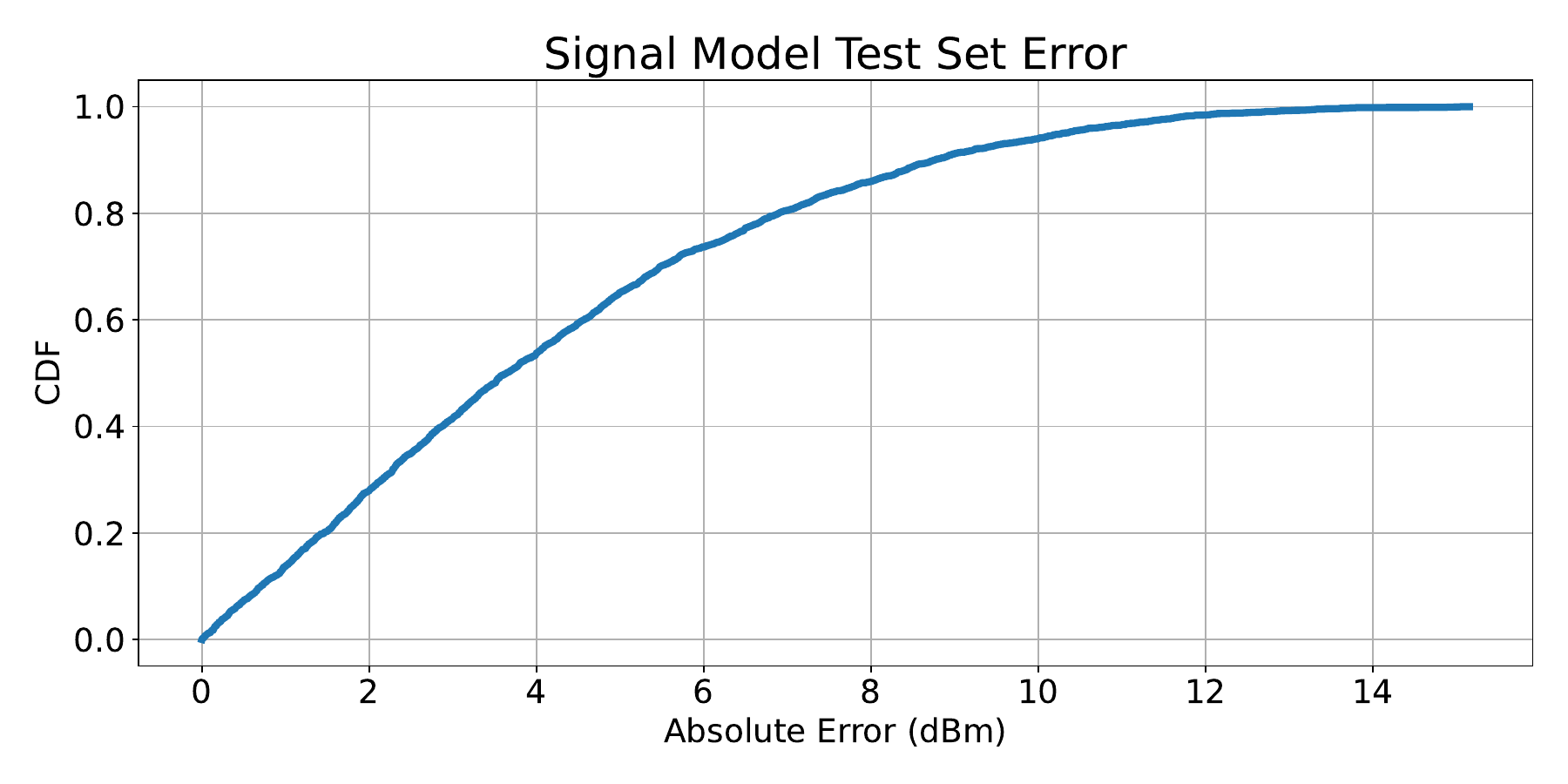}
          \vspace{-0.2in}
    } \hfill
    \subfloat[Predicted vs measured signal for a UAV test trace.]{
        \includegraphics[width=0.35\textwidth,valign=c]{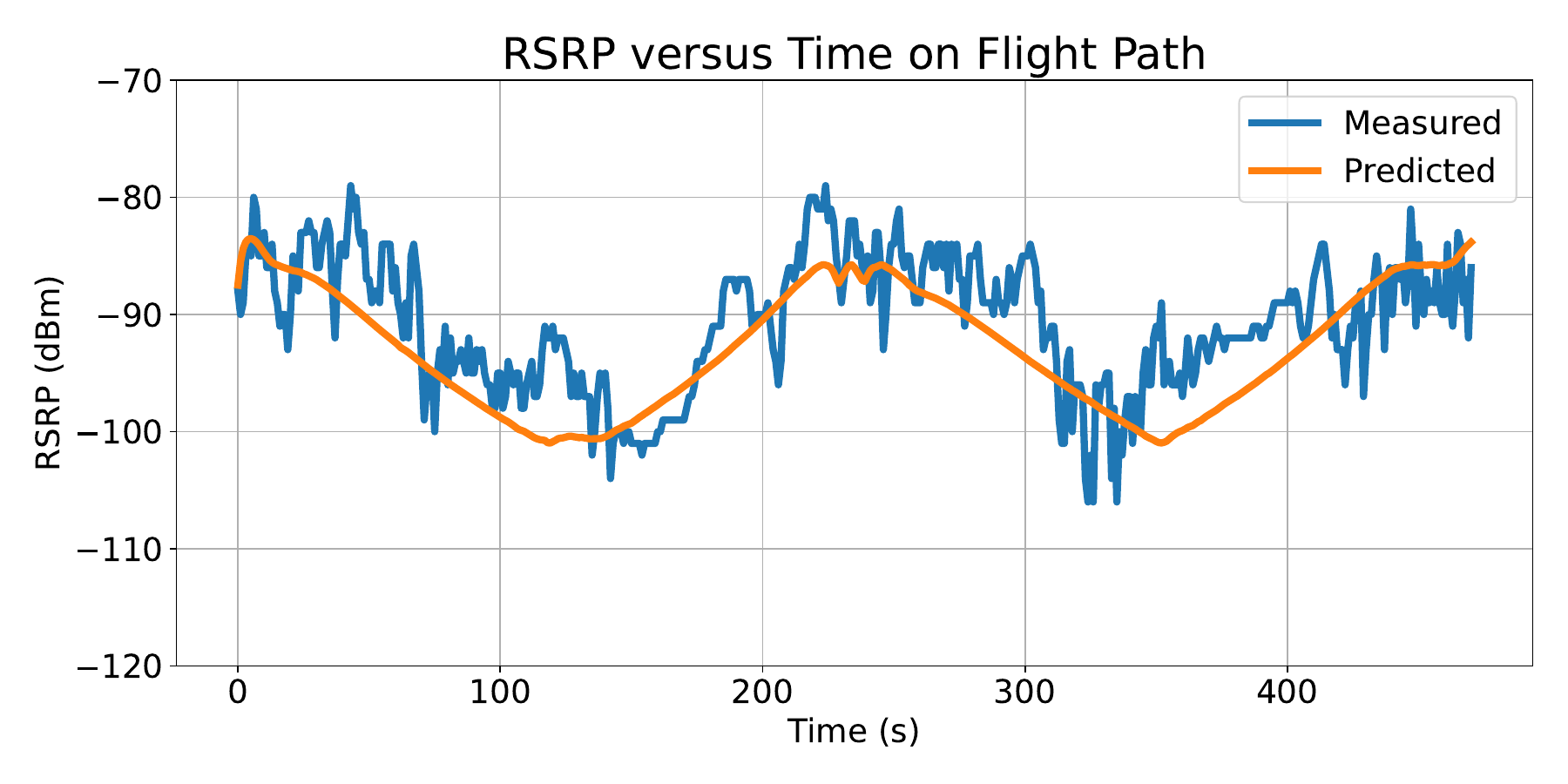}
          \vspace{-0.2in}
    } \hfill
    \subfloat[Test trace flight path. Darker implies higher RSRPs. ]{
        \includegraphics[width=0.17\textwidth,valign=c]{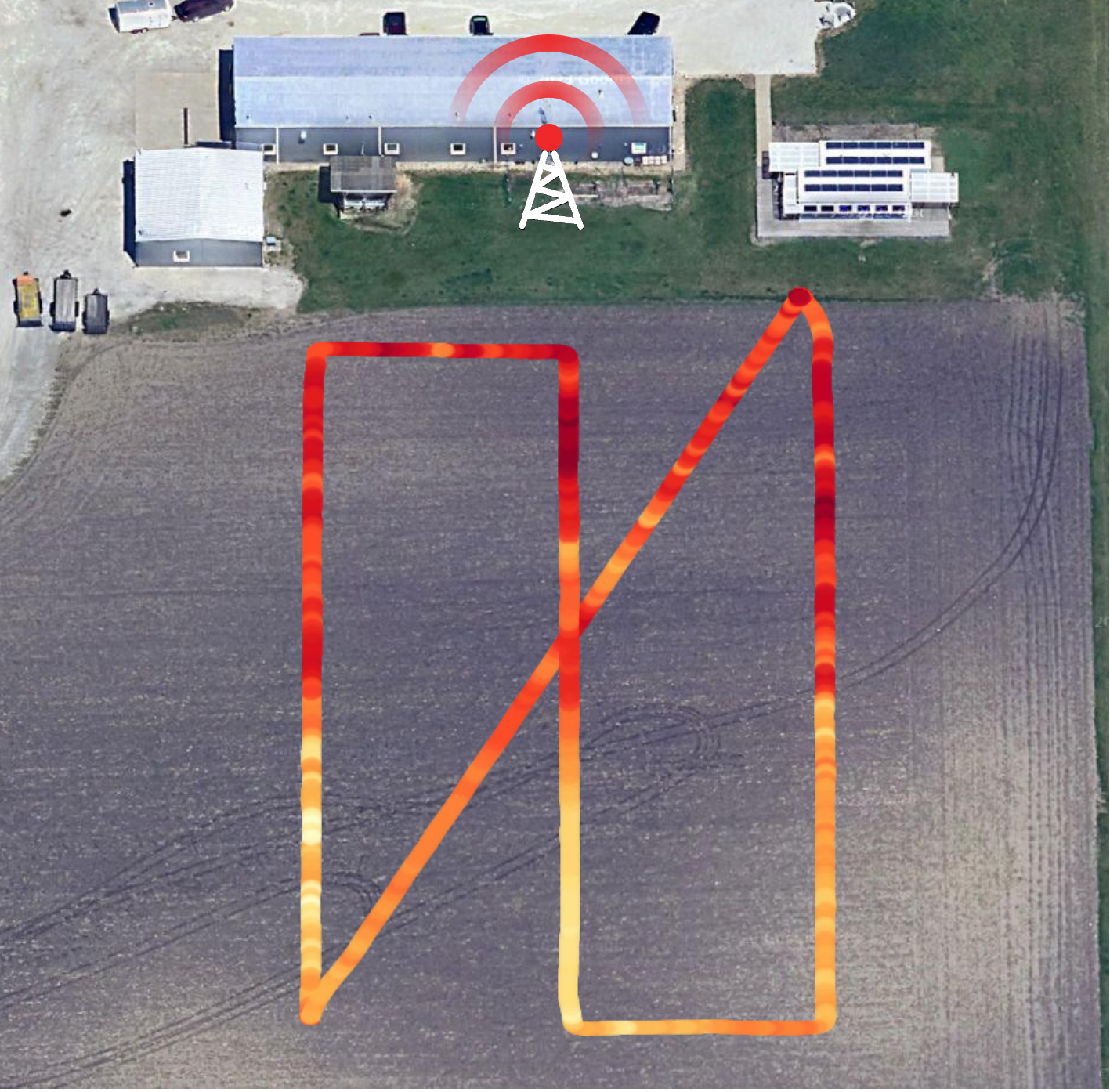}
          \vspace{-0.2in}
    }
    \vspace{-0.2in}
    \caption{\textbf{Evaluating the signal model.} Our model achieves $5.27$ dBm RMSE and $3.65$ dBm median absolute error. }
    \label{fig:signal_model_eval}
    \vspace{-0.2in}
\end{figure*}

\para{Drone Horizontal Motion Dataset: }We collect a new dataset in the following week. We move the drone horizontally for this set of experiments to sample densely in the horizontal plane for different drone heights. Specifically, we fix an altitude $h_{bs} \in \{5, 10, 15\}$m and have the UAV fly a zig-zag survey pattern (see Fig.~\ref{fig:signal_model_eval}(c)) over the field at this altitude. The zig-zag pattern is intended to mimic the path that a real under-crop canopy robot takes when surveying successive crop rows in a field (see Fig.~\ref{fig:phenotyping_route}). In addition to orienting the zig-zag pattern in a north-south direction (as pictured), we also orient it in an east-west direction to capture more spatial variation. We aggregate data points from 9 such flights, resulting in a dataset comprising 2656 usable data points. 

\subsection{Evaluation: Signal Model}\label{sec:eval_model}

We derive the optimal parameters for our signal model (Eqn.~\ref{eqn:signal-model}) given calibration data derived from the vertical motion dataset defined above. The resulting optimal parameters are shown in Table~\ref{tab:model_parameters}. Then, we evaluate our model's accuracy using the horizontal motion dataset described above. Note that the evaluation data was collected one week after the calibration data, so they have no overlap. The results of testing our model on this dataset are shown in Fig.~\ref{fig:signal_model_eval}. As shown, the model achieves a median error of 3.65 dBm. We show an example trajectory of the drone in Fig.~\ref{fig:signal_model_eval}(B-C). The example trajectory demonstrates that our model is able to capture the overall trend very well. There is minor temporal variation, likely due to multipath reflections and noise, that the model doesn't capture which leads to the error.

\begin{table}[h!]
\centering
\vspace{-0.1in}
\caption{Model parameters for peak-season corn.}
\vspace{-0.15in}
\begin{tabular}{cccc}
\toprule
$\bm{\alpha}$ & $\bm{\beta}$ & $\bm{\Gamma}$ & $\bm{G}$ \\
\midrule
 0.501 & 0.185 & 3.741 & -55.420 \\
\bottomrule
\end{tabular}
\label{tab:model_parameters}
\vspace{-0.15in}
\end{table}

\subsection{Evaluation: Predicting Ideal Height}
\begin{figure*}[!ht]
    \subfloat[Comparing RSRP CDF of fixed baseline vs \name.]{
        \includegraphics[width=0.31\linewidth,valign=c]{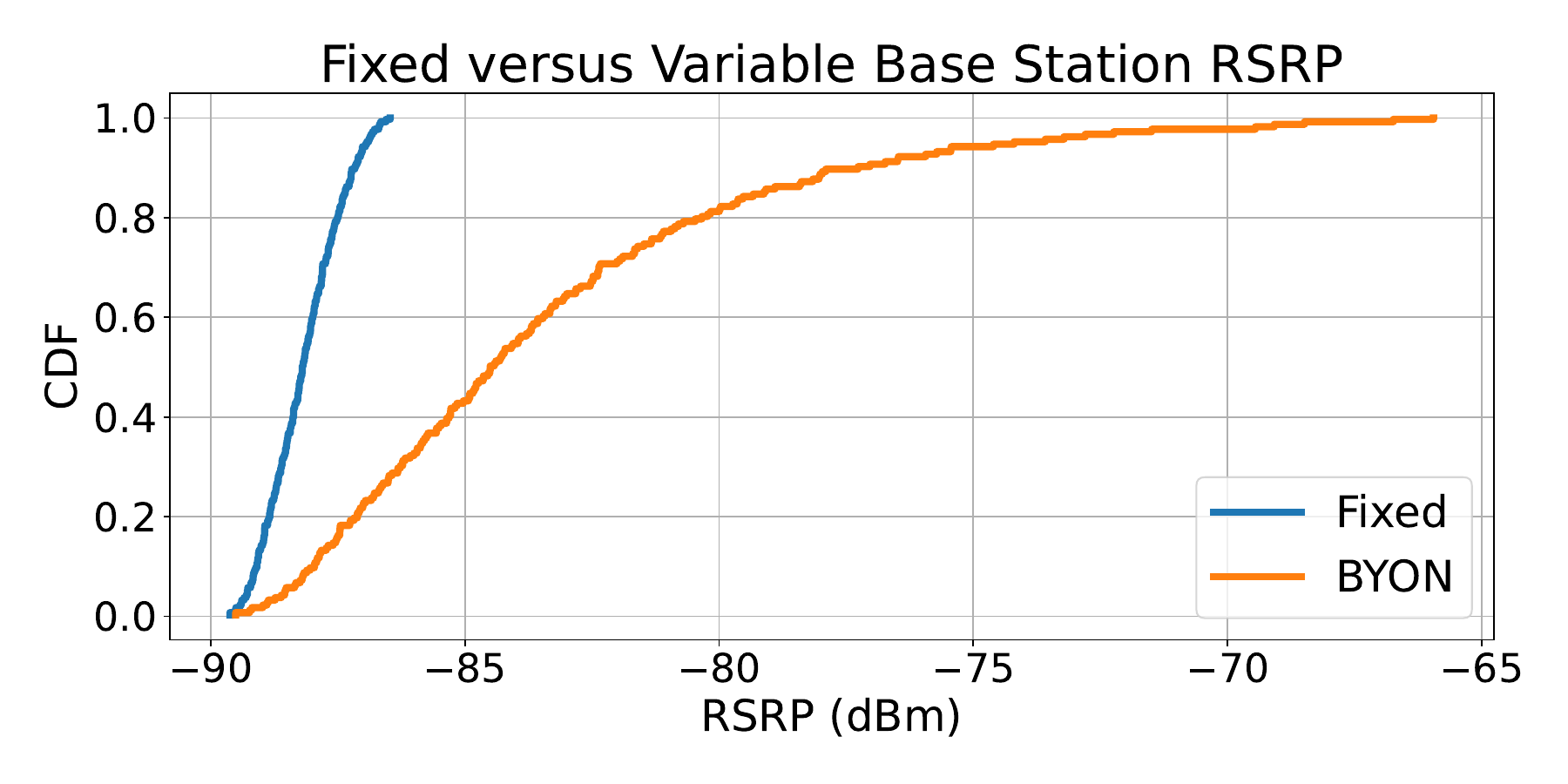}
        \vspace{-0.1in}
        \label{fig:fixed_vs_variable}
    } \hfill
    \subfloat[\name's dynamic base station improves RSRP across our test field by an average of $4.80$ dBm.]{
        \includegraphics[width=0.31\linewidth,valign=c]{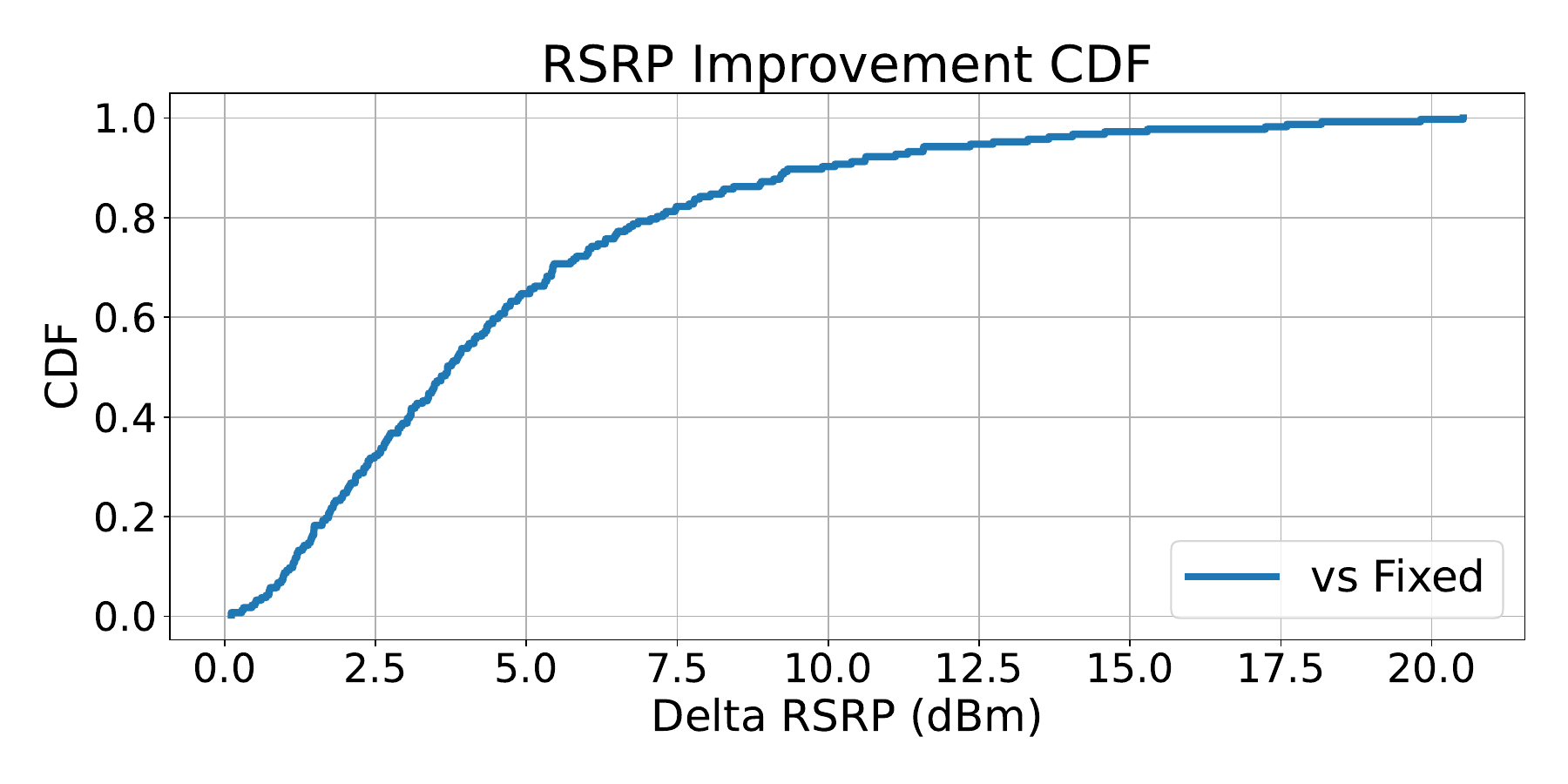}
        \vspace{-0.1in}
        \label{fig:fixed_vs_variable_delta}
    } \hfill
    \subfloat[Multi-client results for 2, 5, 10 devices.]{
        \includegraphics[width=.31\linewidth,valign=c]{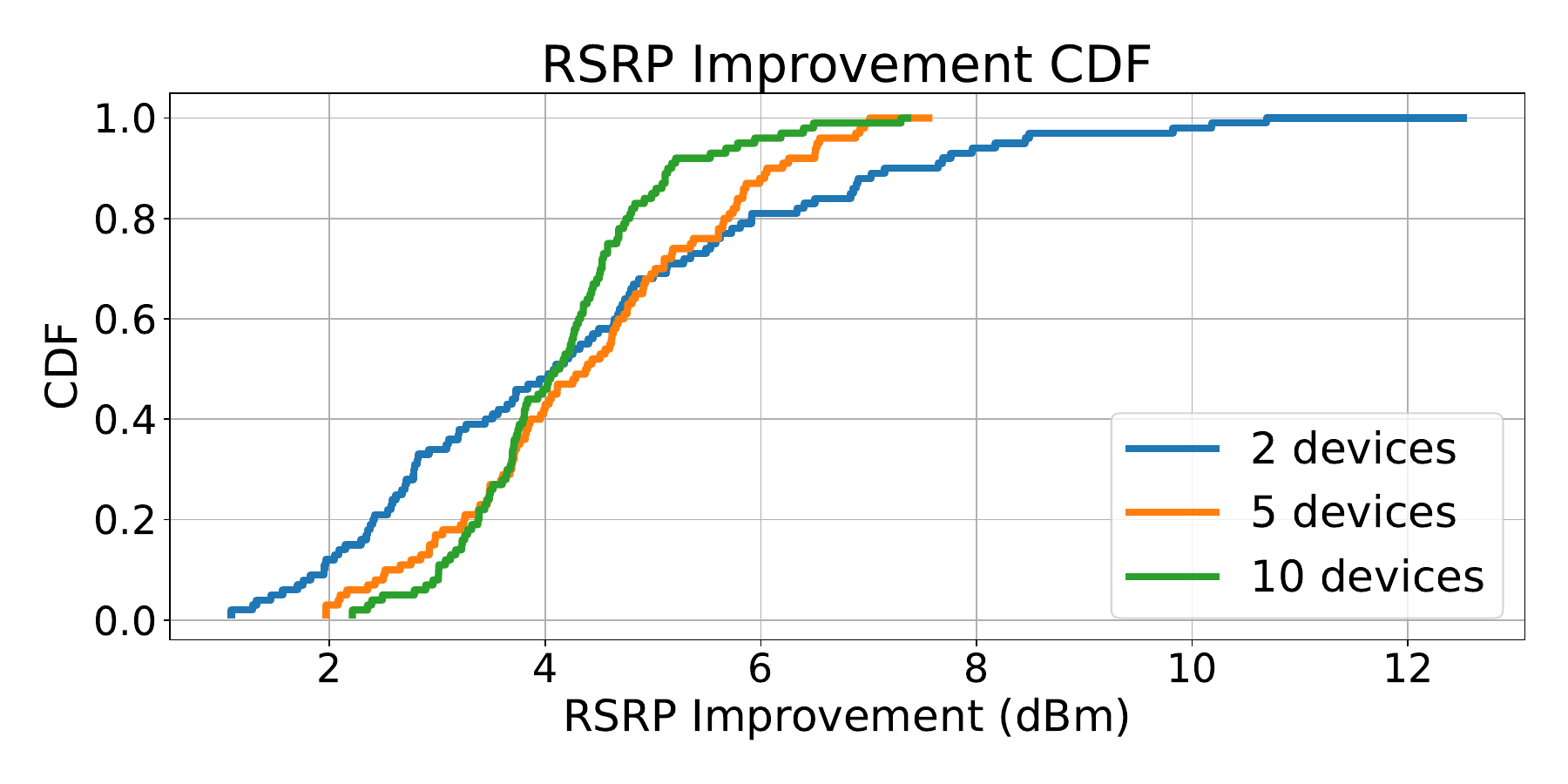}
        \vspace{-0.1in}
        \label{fig:multiple_client}
    }
    \vspace{-0.15in}
    \caption{\textbf{Comparing variable height vs fixed base station. } We compare the RSRP values achieved by the dynamic height base station vs those of fixed height base station at $h_{bs}=5$ m. }
\label{fig:height_eval}
  \vspace{-0.2in}
\end{figure*}

Next, we quantify the benefits of \name's dynamic base station versus a fixed height base station for clients in the field. To do this, we uniformly sample a $20 \times 20$ 2D grid of points in our test field. For ground truth, we use the data collected in Sec.~\ref{sec:dataset}. Specifically, for a given setting of $h_{bs}$, $r$, and $h_{c}$, we need to find the data point that corresponds to this setting and use its value as the ground truth measurement. We interpolate the measurements linearly to fill in missing values.

For each location in our grid, we compare the RSRP (signal strength) obtained using a fixed base station height of $h_{bs}=5m$ and using \name's variable height base station. Our results are shown in Fig.~\ref{fig:fixed_vs_variable} and Fig.~\ref{fig:fixed_vs_variable_delta}.
The baseline achieves a median RSRP of -89.11dBm, our variable-height design achieves a median RSRP of -81.65dBm, which is an improvement of 7.46dBm. Using the signal strength-throughput information shown in Fig.~\ref{fig:vs_signal}, we can see that this improvement would translate to a 28.5\% higher downlink rate for the median device. From Fig.~\ref{fig:fixed_vs_variable_delta}, we also see that \name's average improvement across all devices is 4.8 dBm, with 10\% of the devices getting over 12 dBm signal strength improvement. This experiment demonstrates that \name\ can significantly improve throughput and signal quality for under-canopy applications.

\subsection{Evaluation: Multiple Clients}

Next, we quantify \name's gains under a realistic multiple client scenario. During deployment, several robots can be active at the same time working on different crop rows in parallel, e.g., along the different crop rows in Fig.~\ref{fig:phenotyping_route}. These crop rows are at different distances from the base station. In this scenario, \name\ has the choice of optimizing the throughput for some subset of clients or that of all clients simultaneously. We consider the scenario where \name\ pursues an unbiased policy i.e. optimizes average throughput to all clients. To simulate this scenario, we fix a set of $n \in \{2,5,10\}$ clients. We randomly sample a different crop row within our test field for each client and also randomize the placement of the client within the row. Then for these client locations, we compute the gains for the average RSRP from varying the base station height vs. fixing $h_{BS}=5m$. We repeat this process for $k = 100$ trials and plot the gains as a CDF. We show the results in Fig.~\ref{fig:multiple_client}. We can see that in the scenario of 2,5 and 10 clients, \name\ can achieve an average RSRP improvement of 4.54 dBm, 4.49 dBm, and 4.26 dBm respectively. While under all occasions \name\ was able to greatly improve the signal quality, we do see that such improvement slightly decreases as more devices are added to the field. This agrees with the intuition that when there are more devices, their optimal $h_{BS}$ would be different and it would be harder to optimize the signal quality for all of them.

\section{Case Study: Robot Teleoperation}
\label{sec:application}

\begin{figure*}[!ht] %
  \centering
  \subfloat[Teleoperation has asymmetric uplink and downlink requirements.]{%
    \includegraphics[width=0.22\textwidth, valign=c]{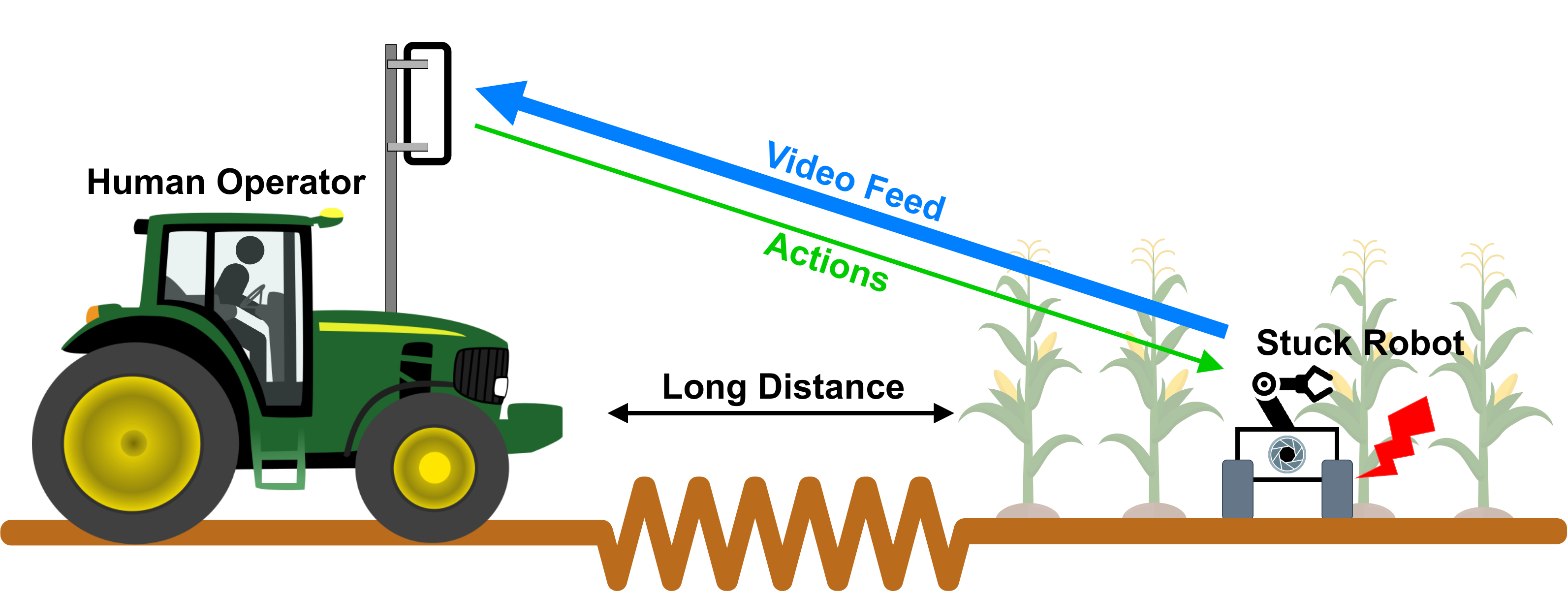} %
    \vspace{-0.2in}
    \label{fig:teleop}
  }
  \subfloat[Video data rate at each compression level.]{%
    \includegraphics[width=0.26\textwidth, valign=c]{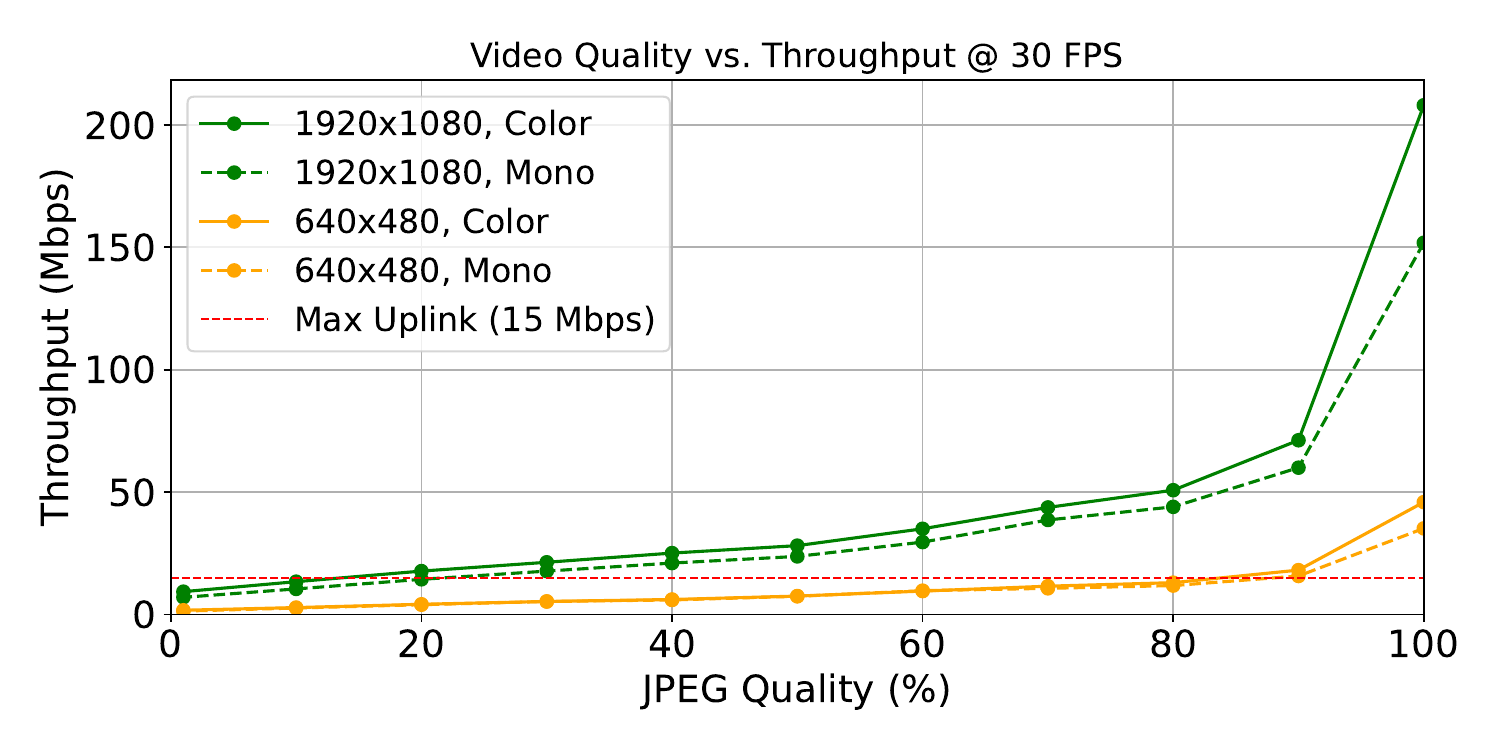} %
    \vspace{-0.2in}
    \label{fig:teleop_bw}
  }
  \hfill
  \subfloat[Video latency at each compression level.]{%
    \includegraphics[width=0.26\textwidth, valign=c]{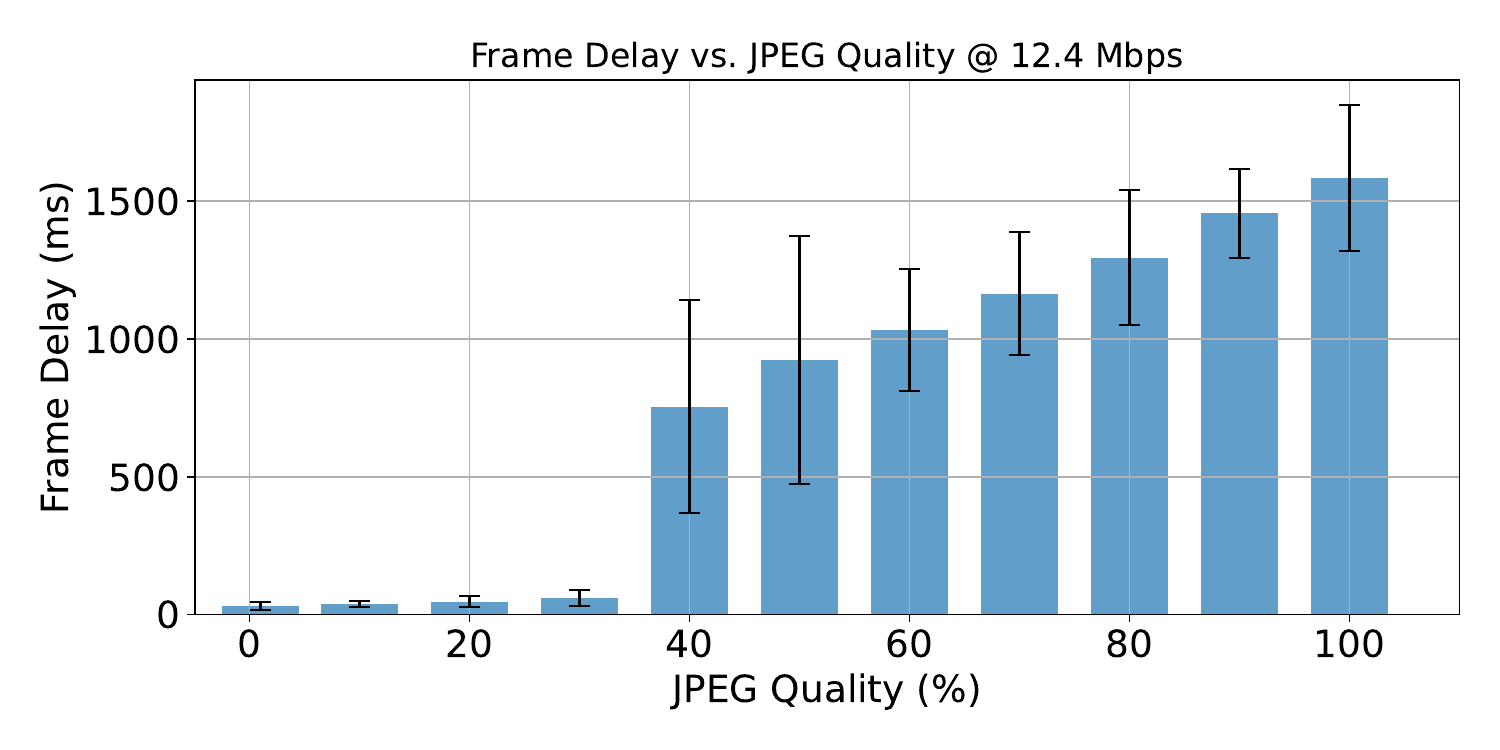} %
    \vspace{-0.2in}
    \label{fig:teleop_delay}
  }
  \subfloat[Frame drop rate versus distance.]{%
    \includegraphics[width=0.22\textwidth, valign=c]{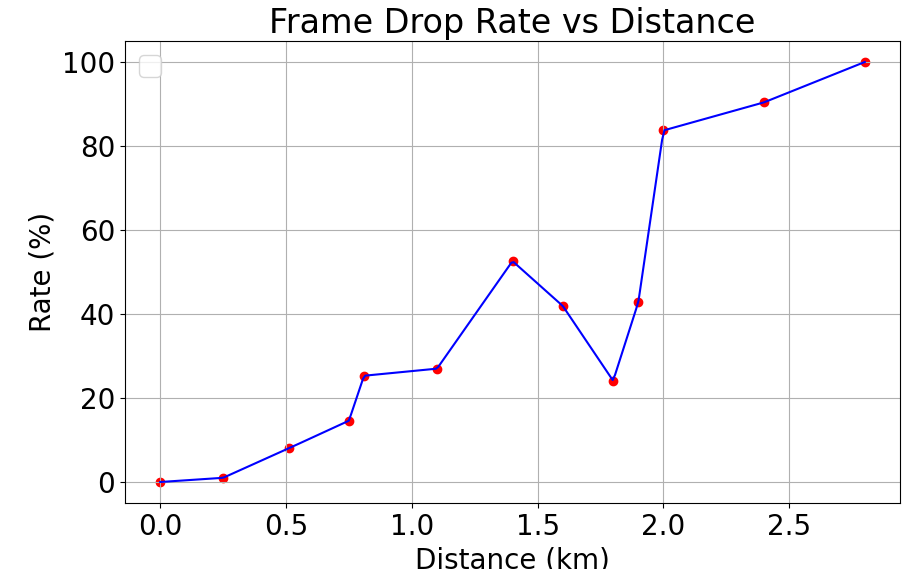}
    \vspace{-0.2in}
    \label{fig:teleop_range}
  }
  \vspace{-0.15in}
  \caption{\textbf{Robot teleoperation.} The operator obtains a video feed from the stuck robot and issues corrective actions. }
  \label{fig:teleoperation}
  \vspace{-0.2in}
\end{figure*}

In this section, we explore the use of BYON in teloperating under-canopy robots. As outlined previously, these small robots fit underneath the crop canopy and are capable of autonomous row following to perform tasks such as plant phenotyping~\cite{phenotyping1,sivakumar_learned_2021,manish_agbug_2021,kim_p-agbot_2022}, cover crop planting~\cite{icover_usda, farmprogress_robot, du_deep-cnn_2022}, and mechanical weeding~\cite{ mcallister_agbots_2020, naio_oz, reiser_development_2019}. However, even with state-of-the-art row-following systems \cite{sivakumar_learned_2021,velasquez_multi-sensor_2022,gasparino_cropnav_2023}, such robots require frequent intervention (e.g., once per 250 m~\cite{sivakumar_learned_2021}). When a robot gets stuck, a human operator needs to get to the robot and manually maneuver it, e.g., using a controller operating over WiFi link to the robot. Therefore, successful operation of these robots is labor-intensive.

We examine the use of BYON to solve this problem. With BYON, the robots and the operator are connected to the same CBRS network. This allows an operator to manage a fleet of robots over long distances from a terminal connected to the base station (see Fig.~\ref{fig:teleop}). When a robot requires intervention, the operator can teleoperate the robot from their tractor or office. This involves remotely issuing actions to the robot while monitoring its video feed.

\para{Throughput Requirements:}
First, we analyze throughput requirements for teleoperation. As depicted in Fig.~\ref{fig:teleop}, we note that teleoperation has asymmetric uplink and downlink throughput requirements. We require a high throughput uplink in order to stream video from the robot. By contrast, the downlink consists of small command messages from the teleoperator. Hence, we are bottle-necked by the limited uplink of the CBRS network. We explore several means of reducing the data rate of the video on the robot. We explore the effect of varying color channel, resolution, and compression levels. Our results are shown in Fig.~\ref{fig:teleop_bw}. The greatest effect stems from resolution and compression levels. We find the network supports 640x480 video under most compression levels.

\para{Latency Requirements:} Next, we analyze latency requirements for teleoperation. Teleoperation favors lower latency because it is easier when feedback is more immediate. To find the relationship between compression levels and latency, we conduct an experiment by varying video compression levels on a uplink that averages 12.4 Mbps. Our results are shown in Fig.~\ref{fig:teleop_delay}. We find increasing compression levels decreases latency, with delay and jitter increasing sharply below a compression threshold. 

\para{Choosing Compression Level:} The preceding experiments suggest we should increase the video compression as much as possible to decrease data rate and latency. However, increasing the compression limits the ability of the operator to understand the video. Hence, there is a tradeoff, and we are interested in finding the minimum video quality that an operator needs to effectively teleoperate a robot. To do this, we have a human attempt to drive the robot through crop rows at various compression levels. We find that human operators can tolerate compression levels as high as 90\%.

\para{Limits of Teleoperability: } Next, we ascertain the range limits of teleoperation. We fix the video compression level to 90\% (3.6 Mbps), adjust the base station (Fig.~\ref{fig:base_station}) power to normal levels (50W) and attempt to teleoperate the robot at distances of up to 2.5 km away (Fig.~\ref{fig:teleop_range}). We observe that as the distance increases, the rate of frame drops in the video increases. Empirically, we find drop rates of up to 30\% to be tolerable. In summary, we are able to teleoperate robots up to 2.0 km away without crops, up to 700 m with early season crops, and up to 200 m with peak season crops. Note that these are all in excess of Wi-Fi's range in open spaces ($\sim 100$ m). Finally, we note that some techniques (e.g. delay compensation \cite{chakraborty_towards_2024}) could further increase the range; we delegate such investigations to future work.

\para{Scalability: } How well does \name\ scale to multiple robots? Our application is bottlenecked by the uplink to the base station \cite{celona_ap11}, which supports up to 50 Mbps of uplink capacity from all clients. If we assume a conservative 20\% JPEG quality video stream consuming 4.2 Mbps (refer to Fig.~\ref{fig:teleop_bw}), we can support up to 10 robots concurrently, depending on their relative location. %

\section{Concluding Discussion}
\label{sec:conclusion}
We present \name -- a new connectivity model for agricultural applications. We highlight the challenge of under-canopy connectivity for both CBRS and satellite signals. We present a height-variable base station design to alleviate performance degradation due to crops. Finally, we demonstrate under-canopy robot teleoperation. We conclude with some remarks:

\para{Deployment Models: }We envision two deployment models for \name. In a community-driven model, a community may share a \name\ setup and use it to serve multiple farms. Typically, both farming and harvest seasons are spread out over a few weeks in a given community. So, such sharing models may be feasible. Second, some robots/equipment companies work on a rental model, where farm equipment or even a planting/harvesting/cover-cropping service is leased to a farmer. In such cases, the equipment/service provider can bring their own \name\ infrastructure. 

\para{Self-Calibration for Different Crops/Weather: } In this work, we considered CBRS signal propagation through dry corn. To adapt \name\ to different crops and weather conditions, it is only necessary to change the attenuation coefficient $\alpha$ accordingly. Note that \name\ knows the signal reading and location of clients, location and height of its base station, and the crop height. Thus, it is possible to adapt the attenuation coefficient $\alpha$ online by inverting the signal model. Thus, \name\ can continually adapt itself accordingly for different crops and weather conditions throughout the season. We leave such exploration to future work.

\definecolor{green}{RGB}{0, 128, 0}
\definecolor{red}{RGB}{255, 0, 0}
\definecolor{mediumyellow}{RGB}{220, 204, 0}

\newcommand{\greencheck}{\textcolor{green}{\checkmark}}
\newcommand{\xmark}{\textcolor{red}{\ding{55}}}

\clearpage 

{ \bibliographystyle{acm}
\bibliography{reference}}

\end{document}